\documentclass[structabstract]{aa}  
\usepackage[dvips]{graphicx}
\usepackage{longtable,lscape}
\usepackage{txfonts}
\begin{document}
   \title{Spectroscopic characterization of 78 DENIS ultracool dwarf 
candidates in the solar neighborhood and the Upper Sco OB association}


   \author{
E. L. Mart\'{\i}n \inst{1, 2}
N. Phan-Bao, \inst{3, 4}
M. Bessell,  \inst{5}   
X. Delfosse, \inst{6} 
T. Forveille, \inst{6} 
A. Magazz\`u,  \inst{7}
C. Reyl\'e,  \inst{8} 
H. Bouy,  \inst{1}
R. Tata  \inst{2}
                    }

\offprints{N.~Phan-Bao}

\institute{Instituto de Astrof\'{\i}sica de Canarias, 
C/ V\'{\i}a L\'actea s/n, E-38200 La Laguna (Tenerife), Spain \\
 \email{ege@iac.es}
\and
 University of Central Florida, Dept. of Physics, 
PO Box 162385, Orlando, FL 32816-2385, USA\\
 \email{tata@physics.ucf.edu}
\and
Department of Physics, HCMIU, Vietnam National University Administrative Building,
Block 6, Linh Trung Ward, Thu Duc District, HCM, Vietnam \\
\email{pbngoc@hcmiu.edu.vn} 
\and
Institute of Astronomy and Astrophysics, Academia Sinica, 
PO Box 23-141, Taipei 106, Taiwan\\
 \email{pbngoc@asiaa.sinica.edu.tw}
\and
Research School of Astronomy and Astrophysics, 
College of Science, Australian National University,
ACT 2611, Australia\\
 \email{bessell@mso.anu.edu.au}
\and
Laboratoire d'Astrophysique de Grenoble, Universit\'e 
J. Fourier, BP 53, 38041 Grenoble, France\\
 \email{[forveille,delfosse]@obs.ujf-grenoble.fr}
\and
Telescopio Nazionale Galileo, 
Rambla Jos\'e Ana Fern\'andez P\'erez 7 E-38712 Bre\~na Baja, 
Spain\\
 \email{magazzu@tng.iac.es}
\and
Observatoire de Besan\c{c}on, Institut UTINAM, University of Franche-Comt\'e, 
CNRS-UMR 6213, BP 1615, 25010 Besan\c{c}on Cedex, France\\
 \email{ celine@obs-besancon.fr}
  }

   \date{Received ; accepted }

 
  \abstract
   {}
   {Report Low-resolution optical spectroscopic observations for 78 very low-mass star and brown dwarf 
candidates that have been photometrically selected using the DENIS survey point 
source catalogue.}
   {Spectral types are derived for them using measurements of the   
PC3 index. They range from M6 to L4. 
H$\alpha$ emission and NaI subordinate doublet 
(818.3 nm and 819.9 nm) equivalent widths are measured in the spectra. 
Spectroscopic indices of TiO, VO, CrH and FeH 
molecular features are also reported. }
   {A rule-of-thumb criterion to select young very low-mass objects using 
the NaI doublet equivalent width is given. It is used to confirm 
seven new members of the Upper Sco OB association and two new members 
of the R Cr-A star-forming region. Four of our field objects are also classified 
as very young, but are not members of any known nearby young association. 
The frequency of lower-gravity young objects in our field ultracool sample is 8.5\% . 
Our results provide the first spectroscopic classification for 
38 ultracool dwarfs in the solar vicinity with spectrophotometric distances 
in the range 17 pc to 65 pc (3 of them are new L dwarfs within 20 pc).}
   {}


\authorrunning{Mart\'{\i}n et al.}
\titlerunning{78 DENIS ultracool dwarf candidates}

   \maketitle
%

\section{Introduction}

   A complete census of the solar neighborhood is needed for many 
different purposes. To mention just a few examples: the 
understanding of the star formation history of the Milky Way;  
the identification of bright benchmark objects of different 
spectral types and evolutionary history, and the search for extrasolar planets. 
Henry et al. (2002) estimated that 
more than half of the stellar systems within 25 pc are 
still unknown, and most of them are thought to be ultracool dwarfs 
(UDs; defined as those with spectral types M6--M9, 
L and T; Mart\'\i n et al. 1996, 1997, 1999; 
Kirkpatrick et al. 1999, 2000), which include a mixture of very low-mass 
stars  and substellar-mass brown dwarfs.  Indeed, 
many UDs within 25 pc have been found in the last few years 
(Phan-Bao et al. 2008; Reid et al. 2008). 

Large optical/infrared surveys have made a significant impact in the 
identification of our coolest neighbours. The Deep Near Infrared Survey 
of the Southern Sky (DENIS; Epchtein et al. 1997) has enabled the discovery of many UDs  
in the solar vicinity (Delfosse et al. 1997, 2001; 
Crifo et al. 2005; Phan-Bao et al. 2001, 2003, 2008; Kendall et al. 2004) 
and prompted the development of a new spectral 
class, the L dwarfs (Mart\'\i n et al. 1997, 1999; Delfosse et al. 1999), 
which is characterized by the condensation of dust grains in the atmospheres 
(Allard et al. 2001; Marley et al. 2002; Tsuji 2005). 

The Sloan Digital Sky Survey (SDSS; York et al. 2000) and the 
Two Micron All Sky Survey (2MASS; Skrutskie et al. 1997) have also brought about 
the discoveries of 
many UDs  (Kirkpatrick et al. 1999, 2000; Fan et al. 2000; Knapp et al. 2004, Cruz et al. 2007, 
Reid et al. 2008) and have provided the identification of the coolest spectral class, 
the T dwarfs, characterized by the presence of methane bands in 
the near-infrared spectrum (Burgasser et al. 2006; Kirkpatrick 2005). 
The new generation large infrared surveys CFHTLS and UKIDSS are starting  
to identify even cooler dwarfs (Warren et al. 2007), 
the so-called Y class, for  which the distinguishing characteristic may be the emergence 
of ammonia molecular bands in the near infrared. 
Two objects with possible ammonia absorption have recently been 
identified (Delorme et al. 2008), but their classification as Y dwarfs remains controversial 
because they are very similar to late T dwarfs (Burningham et al. 2008).

The characterization of the nearby ultracool population continues to be a 
goal of recent papers. Besides those mentioned above, it is worthwhile to mention 
a few more. Kendall et al. (2007) presented twenty-one southern ultracool dwarfs 
(M7--L5.5) selected from 2MASS and SuperCOSMOS point 
source databases according to their colors and proper motions, and confirmed via 
low-resolution near-infrared spectrosocopy. Costa et al. (2006) and Henry et al. (2006) 
reported trigonometric parallaxes for several UDs, including the closest known L dwarf. 
Jameson et al. (2008) have provided proper motion measurements for over a 
hundred L and T dwarfs. 

In this paper we present low-resolution optical spectra of 78 UD 
candidates selected from the DENIS point source catalog using 
photometric color criteria. 
For a subsample of them (50 objects with galactic latitude between 30 and 15 degrees), 
the Maximum Reduced Proper Motion (MRPM) method 
(Phan-Bao et al. 2003, 2008) is used to reject giants. We also present spectra of a few DENIS UD 
candidates in the general area of the Upper Sco OB associations.


\section{SAMPLE SELECTION AND SPECTROSCOPIC OBSERVATIONS}

Most (71 out of 78) of our sample comes from a systematic search of 10,000 
square degrees of the DENIS database (available at the Paris Data 
Analysis Center, PDAC) for potential UD members of the solar 
neighbourhood that are redder than $I-J \geq 3.0$ and have galactic 
latitudes $|b_{II}|~{\geq}~15\degr$ (Delfosse et al. 2003). 
21 of them have galactic latitudes $|b_{II}|>30$, and the other  
50 were extracted from a selection with 
a galactic latitude criterion less restricting ($|b_{II}|>15$) and for which the 
MRPM method was used to discriminate mearby ultracool 
dwarfs from distant red giants (Phan-Bao et al. 2003, 2008). Proper motions were
measured using Aladin and the Digital 
Sky Survey\footnote{http://archive.stsci.edu/cgi-bin/dss\_plate\_finder} (DSS). 
They will be given in another paper that will deal with the analysis of the 
kinematics of the UDs. 
7 of our targets come from a search for low-mass members of the Upper Sco 
OB association (Mart\'\i n, Delfosse \& Guieu 2004). We covered 
60 square degrees looking for objects redder than $I-J \geq 2.3$. 
The names, coordinates and photometry of all targets are provided in Table 1.

Spectroscopy of DENIS candidates presented herein 
was obtained in several observing runs using different 
telescopes. Following a chronological order, 
we started on March 2000 with ALFOSC on the Nordic Optical Telescope 
in La Palma. The grism number 5 provided a 
dispersion of 3.1 \AA ~ per pixel. Our second observing run was in  
September 2000 with 
the red arm of the ISIS spectrograph mounted on the 4.2-m William 
Herschel telescope in La Palma. The grating R158R provided a 
dispersion of 2.9 \AA ~ per pixel. 
 The same telescope and instrumental 
configuration were again used on December 2006. However, a different 
CCD was installed, resulting in a dispersion  of  1.6 \AA ~ per pixel. 
A slit of 1 arcsec gave a spectral resolution of 6.5 \AA . 

On December 2000 and November 2003, we used the ESO NTT 
with the EMMI intrument in its Red Imaging and Low-Dispersion mode
(RILD). In this mode the dispersion is 2.8 \AA~pixel$^{\rm -1}$, 
and the effective wavelength range is 520 to 950 nm. 
The spectrophotometric standards, LTT~2415 and Feige~110 were chosen
from the ESO list. All reduction was performed within MIDAS.
We selected the 1\arcsec~ slit, which corresponds to a spectral resolution
of 10.4~ \AA~. 

Spectroscopic data of more DENIS 
candidates were collected on August 2002 with 
the FORS2 spectrograph mounted on the 8-m Very Large Telescope  
in Paranal. The grating 600I provided a 
dispersion of 1.3 \AA ~ per pixel. 
Additional DENIS objects were observed at the 2.3-m telescope 
of the Sidings Springs Observatory (SSO) in Australia 
between 22 and 28 June 2006.  
The Double Beam Spectrograph with 
gratings 158R and 316R was used providing dispersions  
of 3.7 \AA ~, and  1.87 \AA ~respectively. 
The last observing run included in this paper took place on 17 July 2007 
at the Blanco 4-m telescope in the Cerro Tololo Interamerican Observatory 
(CTIO).  
The spectroscopic observing log that summarizes 
all these observations is provided in Table 2.  

All the spectra were reduced following standard procedures within 
the IRAF environment (bias and flatfield correction, wavelength 
calibration using a CuNeAr lamp, and flux calibration using standards), 
except the SSO spectra which were reduced in the FIGARO environment.

\section{Spectral Types}

Low-resolution CCD spectra have been used to define the 
spectral classification of late-M and early-L dwarfs 
(Kirkpatrick et al. 1999; Mart\'{\i}n et al. 1996, 1999). The PC3 index defined by 
the latter authors has been used in several papers to determine spectral types 
(Crifo et al. 2005; Mart\'{\i}n et al. 2004, 2006; Phan-Bao \& Bessell 2006, 2008; 
Reyl\'e et al. 2006). 
Comparisons with other methods of spectral type determination have found 
that the results are consistent. We have measured the PC3 index in our spectra, 
and we have determined spectral subclasses following the relationships given 
by Mart\'{\i}n et al. (1999). The results are given in Table 3. 

We obtained spectra for seven UDs in common with Mart\'{\i}n et al. (1999). The 
spectral types obtained from measurement of the PC3 index in our spectra 
are in very good agreement with their values. Therefore the spectral classification 
adopted in this paper is tied to that of Mart\'{\i}n et al. (1999), which is consistent 
within one spectral subclass with that of Kirkpatrick et al. (1999) for dwarfs earlier than 
L4. All the spectral types were checked by visual inspection and compared 
to standards from Mart\'{\i}n et al. (1999). 
Full spectra of some of the latest type objects in our sample are shown in Figure 1. 

We found that 34 of our targets already had spectral types in the literature 
(see references in the caption to Table 3). The largest overlap of objects 
with spectral types estimated independently  is with Reid et al. (2008). 
In Figure 2, we show the comparison between our spectral types and those 
in the literature. Generally, there is a fairly good agreement; most of 
the dwarfs have the same spectral type within $\pm$1 subclass. We find 
spectral subclasses consistent  
with Reid et al. 2008 within 1 spectral subclass for all the 12 UDs in common. 

For 5 objects we obtain spectral subclasses that deviate by more than 1 subclass 
with respect to those published in other papers. Three of them, namely  
DENIS J0314352$-$462341, DENIS J0301488$-$590302 and DENIS J1216121$-$125731 had 
spectral types estimated from $I-J$ colors (Bouy et al. 2003). Our spectral type determination 
supersedes theirs, and underscores the limitations of using color-spectral class 
relations for ultracool dwarfs. The other two objects are 
DENIS J0357290$-$441731 and DENIS J1004283$-$114648. Both of them are binaries, 
and therefore our spectra are composites of 2 components of different spectral class. 
The spectral types of the resolved components given by Mart\'{\i}n et al. (2006) 
should be more accurate than ours. 

\section{Surface gravity} 

Spectroscopic surface gravity estimates have been used to identify 
giants and young brown dwarfs among samples of ultracool candidates 
(Mart\'{\i}n et al. 1996; Luhman et al. 1998; Gorlova et al. 2003; 
Mart\'{\i}n \& Zapatero Osorio 2003; McGovern et al. 2004; Allers et al. 2006). 
Atomic gravity-sensitive features present in our spectra include the  
KI resonance doublet at 766.5 nm and 766.9 nm, and the NaI subordinate doublet 
at 818.3 nm and 819.9 nm. 
We have used the NaI doublet because it is located in a region of stronger pseudocontinuum 
emission and it is less affected by telluric lines than the KI doublet. 
The measured equivalent widths of the NaI doublet are given in Table 4. 
Due to the low spectral resolution of most of our spectra, we chose to measure 
the combined equivalent width of the two lines. 

Figure 3 illustrates the 
dependance between NaI equivalent widths and spectral class in our sample. 
There is a lot of scatter in the figure and no clear trend between NaI and spectral class. 
Part of this spread in equivalent widths could be due to the different instruments 
used in this work and to measurement uncertainties. For example, VB10 was observed 
at SSO with a spectral resolution of 8.5 \AA , and at CTIO with a spectral resolution 
of 7.5 \AA . The two spectra are overplotted in Figure 4. Note that the NaI doublet 
appears to be wider in the lower resolution spectrum because of blending effects with 
other absorption features. To measure consistent equivalent widths in spectra of 
different resolutions we established as a rule that the pseudo-continuum region was 
between 823~nm and 827~nm, 
and we integrated the line from 817.5~nm to 821.0~nm. 
The measurements were done 
manually with the IRAF task splot, and the exact integrations limits and continuum 
levels were judged individually for each spectrum, keeping the rule as a general 
guideline, but modifying it slightly as required by the shape of the observed spectrum. 
For our two spectra of VB10, we derived equivalent widths of 7.3$\pm$0.6 \AA , 
and 6.2$\pm$0.3 \AA , 
respectively. The lower value corresponds to the higher resolution spectrum. 
We conclude that the difference in equivalent width due to observing the same 
object with different instruments can be larger than the uncertainty 
in the equivalent width measurement. 

From Figure 3, we infer that the objects with the weakest NaI 
are likely to have low surface gravity. As a  rule of thumb we can state that {\it  
any object of spectral class between M6 and L4, and with a  
NaI (818.3,819.9 nm) doublet detectable but weaker than field counterparts observed with the 
same spectral resolution, is likely to have a low surface gravity and 
consequently a very young age (younger than the Pleiades cluster, i.e. 100 
Myr) and a substellar mass}. 
Note that the equivalent width determination may depend on the spectral resolution 
of the observations, and thus it is important to compare the young UD candidates with objects 
observed with similar spectral resolution. 

As shown in Figure 3, all of the 7 targets in the Upper Sco OB association (solid hexagons) 
have weak NaI as expected for young brown dwarfs. Putting this result together with the 
28 very low-mass objects confirmed by  Mart\'\i n et al. 2004, brings the total number of 
DENIS discovered Upper Sco members to 35.

Two of our candidates have relatively early M type and no detectable NaI doublet. 
They are classified as giants in Table 4. Such a low contamination by giants 
is consistent with previous results and it is expected because the faint 
magnitudes of our candidates would place them (if they were giants) at distances 
over 300 parsecs from the Galactic disk.

The following field objects have weak NaI, indicative of lower surface gravity: 
DENIS-P J0141582$-$463358 (L0), 
which was suggested to be a young brown dwarf or planetary mass 
object by Kirkpatrick et al. (2006); DENIS-P J0006579$-$643654 (L0); 
DENIS-P J0443373$+$000205 (M9.5); 
DENIS-P J1703356$-$771520 (M9); 
DENIS-P J1901391$-$370017 (M8) and DENIS-P J1935560$-$284634 (M9.5). 
Figures 5 and 6 show comparisons of the spectra of some of these objects 
compared with dwarfs of the same spectral class observed with the same intrumental setup. 
In addition to the weak NaI doublet, the lower-gravity objects also have stronger VO bands 
and weaker FeH bands than their higher-gravity counterparts. 
Figure 7 illustrates this effect; the low-gravity objects display higher values 
of the VO/FeH molecular band ratio than the rest of the dwarfs in our sample.   
Figure 8 displays the full spectra of five of our low-gravity field objects, and 
of one of our Upper Sco BDs. 

DENIS-P J1901391$-$370017 is located in the region of the Corona Australis (R-CrA) 
molecular cloud complex. There are 29 young objects listed in SIMBAD 
in an area of 2 arcmin around the DENIS 
source position, and one infrared source at only 3 arcsec. 
In fact the infrared source reported by Wilking et al. (1997) is the same 
as DENIS-P J1901391$-$370017 because the apparent magnitudes are consistent. 
The DENIS object has a low surface gravity and M8 spectral type, and it 
could be the second brown dwarf discovered with DENIS 
in the R-CrA star-forming region after DENIS-P J1859509$-$370632 (Bouy et al. 2004). 
On the other hand, DENIS-P J1935560$-$284634 (M9.5) is near this region, 
and could also be related to the R-CrA star-formation region. This object 
could be the third substellar-mass member detected by DENIS in this region and 
it deserves further attention. 

DENIS-P J0006579$-$643654; DENIS-P J0443373$+$000205 and DENIS-P J1703356$-$771520 
appear to be examples of a very young 
isolated brown dwarf or planetary-mass object that are not obviously associated 
with any known star-forming region. However, 
we note that DENIS J0608528$-$275358, which was identified 
by Cruz et al. (2003) as a young object because of enhanced VO absorption, does not 
have a remarkable weak NaI doublet, and thus its young brown dwarf status remains 
unconfirmed.

A subset of our targets have already been recognized as high proper motion objects. 
DENIS-P J0031192$-$384035; DENIS-P J0050244$-$153818;  
DENIS-P J0227102$-$162446; DENIS-P J0921141$-$210445; 
DENIS-P J1019245$-$270717;  were previously known to be 
high proper motion objects (Deacon, Hambly \& Cooke 2005). 
They are listed in  Simbad with the names of 
2MASS J00311925$-$3840356 or SIPSJ0031$-$3840; 2MASS J00502444$-$1538184 or 
SIPS J0050$-$1538; 2MASS J02271036$-$1624479 or SIPS J0227$-$1624; 
2MASS J09211410$-$2104446 or SIPS J0921$-$2104; 
2MASS J10192447$-$2707171 or SIPS J1019$-$2707, respectively. 
We confirm all of them as higher-gravity old nearby UDs on the basis of their 
strong NaI doublet and 
late spectral type. Their spectrophotometric distances, together with those of 
all other old UDs in our sample, are given in Table 5. We used our adopted spectral types, the 
photometry given in Table 1 and the
absolute J-band magnitude estimated from the absolute
magnitude vs. $I-J$ color relationship given in Phan-Bao et al (2008).
This relationship is not valid for young BDs, and thus we do not give spectrophotometric distances 
from them.   

The fraction of young BDs identified in our field sample (not including Upper Sco) 
is 6/71 (8.5\%). 
Further work to observe these young BDs with higher spectral resolution is needed in order 
to detect lithium, an indicator of youth and substellar mass for UDs (Magazz\`u et al. 
1993), and to compare with theoretical models in order to derive surface gravities.

{\bf \section{A search for dusty disks in young BD candidates}}

We searched the {\it Spitzer} archive for complementary mid-IR data of our young 
BD candidates. 
Three targets (DENIS-P~J0141582$-$463358, J1611296$-$190029 and J1901391$-$370017) 
have been observed with IRAC 
and MIPS. 
Table~6 gives the details of the observations. 
We retrieved the pipeline processed images and extracted the photometry using 
standard PSF photometry routines within the Interactive Data Language (IDL). 
Table~7 shows the photometry of the three sources.  
DENIS-P~J1611296$-$190029 and J1901391$-$370017 do not have any counterpart in the MIPS1 image. 
We derive upper limits by adding a scaled PSF at the expected position of the target until 
the 3-$\sigma$ detection algorithm finds it. 
Uncertainties, including intrumental, calibration and measurement errors, 
are estimated to add up to 10\%. 
Figure 8 shows the spectral energy distribution of the three young BDs with Spitzer data. 
{\bf Comparison with known dwarfs of similar spectral class does not reveal any significant infrared excess
in the three objects as seen in young M dwarfs (e.g., Young et al. 2004) and BDs (e.g., Riaz et al. 2006). 
We therefore conclude that there is no evidence for dusty disks
in these young BD candidates with our current data. }
 
\section{Chromospheric activity} 

H$\alpha$ emission equivalent widths have been determined in our spectra using the 
line integration option in the splot IRAF task. Error bars were assessed object by 
object by repeated measurements using visual judgement of the continuum level 
and the line integration limits. The equivalent width values or upper limits 
are given in Table 4. 

Gizis et al. (2000) reported low-resolution optical spectra for 
53 M7--M9 dwarfs and 7 L dwarfs selected from the 2MASS survey. 
They found that all of their M7--M8 dwarfs displayed 
H$\alpha$ emission, but the frequency of  H$\alpha$ emitters dropped abruptly for cooler 
dwarfs. Similar results were reported by West et al. (2004) in spectroscopic follow-up of 
a large photometrically selected sample from the SDSS. 
In a sample of 152 late-M and L dwarfs, 
Schmidt et al. (2007) found a slightly lower frequency of H$\alpha$ emitters 
among the M8 dwarfs and confirmed the decline of H$\alpha$ emission in L dwarfs. 
Our data indicates a frequency of H$\alpha$ emitters that is consistent with 
that of Schmidt et al. (2007) in the range M8--M9, but it drops faster in the L dwarfs. 
We do not attach high significance to our results because 
it is possible that our lower fraction of H$\alpha$ emitters in the L dwarfs is due 
to the low spectral resolution and modest signal-to-noise ratio of our spectra. 
For example, we observed DENIS J1004283$-$114648 with the NOT and the VLT (Table 2). 
H$\alpha$ emission 
is detected only in the VLT spectrum, which has a resolution 10 times better 
than the NOT spectrum.

In Figure 10, we show the dependance of  H$\alpha$ emission equivalent width 
with respect to spectral class in our sample. The upper envelope of 
chromospheric H$\alpha$ emission given by Barrado y Navascu\'es \& Mart\'{\i}n (2003) 
is shown as a dotted line. 
Only one object lies above this threshold and thus it is a strong candidate 
to harbour an active accretion disk. This object belongs to the Upper Sco sample. Our finding 
of 1 accretor among 7 members in Upper Sco is consistent with the fraction of 
accretors (5/28)  
reported by Mart\'\i n et al. (2004) using the  same criterion.  

All of our 7 objects in Upper Sco and 4 out of 6 of our field lower-gravity objects have 
detected H$\alpha$ emission. 
The frequency H$\alpha$ emitters is higher among the very young objects than 
for the rest of our sample, but 
H$\alpha$ emission is not observed in all young VLM objects. 
As a rule of thumb we can state that 
{\it  H$\alpha$ emission may be an indicator of youth,  
but its detection is not required for an object with a spectral class between 
M6 and L4 to be classified as young.}

We did not detect any obvious flares in our observations. 
Schmidt et al. (2007) estimated a flare cycle of 5\% for late-M dwarfs and of 
2\% for L dwarfs. 
Those numbers may need to be revised slightly downwards. We plan to make a comprenhensive 
study of the flare statistics in UDs in a future paper.

\section{Candidate wide binaries}

Using Simbad, we checked for objects within 2 arcminutes of our targets. 
DENIS J0000286$-$124514 has an X-ray source named 1RXS J000025.0$-$124519. 
No additional information is available on this X-ray source, so it is not 
possible to assess the probability that the DENIS source and the X-ray source 
are related. 

DENIS J1115297$-$242934 has a star named TYC 6653-245-1 with B=11.7 and V=11.1. 
The proper motion of this star is -83.7 and 41.3 mas~yr$^{-1}$ in RA and Dec., respectively. 
According to the NOMAD catalog the proper motion of DENIS J1115297$-$242934 is 
14.0 and -138.0 mas~yr$^{-1}$ in RA and Dec., respectively. Hence, the proper motions 
of the two sources are not consistent with a physical connection.


\begin{acknowledgements}
E.L.M.  acknowledges financial support from NSF grant AST 0440520 
and Spanish MEC grant AYA 2007-67458. 
N.P.-B. has been aided in this  work by a Henri Chretien
International Research Grant administered by the American Astronomical Society. 
X.D. and T.F. acknowledge financial support from the 
"Programme National de Physique Stellaire'' (PNPS)
 of CNRS/INSU, France. 
We thank the referee, John Gizis, for his useful comments on our manuscript.
DENIS is the result of a joint effort involving human and financial
contributions of several Institutes mostly located in Europe. It has
been supported financially mainly by the French Institut National des
Sciences de l'Univers, CNRS, and French Education Ministry, the
European Southern Observatory, the State of Baden-Wuerttemberg, and
the European Commission under networks of the SCIENCE and Human
Capital and Mobility programs, the Landessternwarte, Heidelberg and
Institut d'Astrophysique de Paris.
\end{acknowledgements}

\clearpage
\onecolumn

\begin{figure} 
\vskip 1in
\hskip -0.25in
\centerline{\includegraphics[width=5in,angle=0]{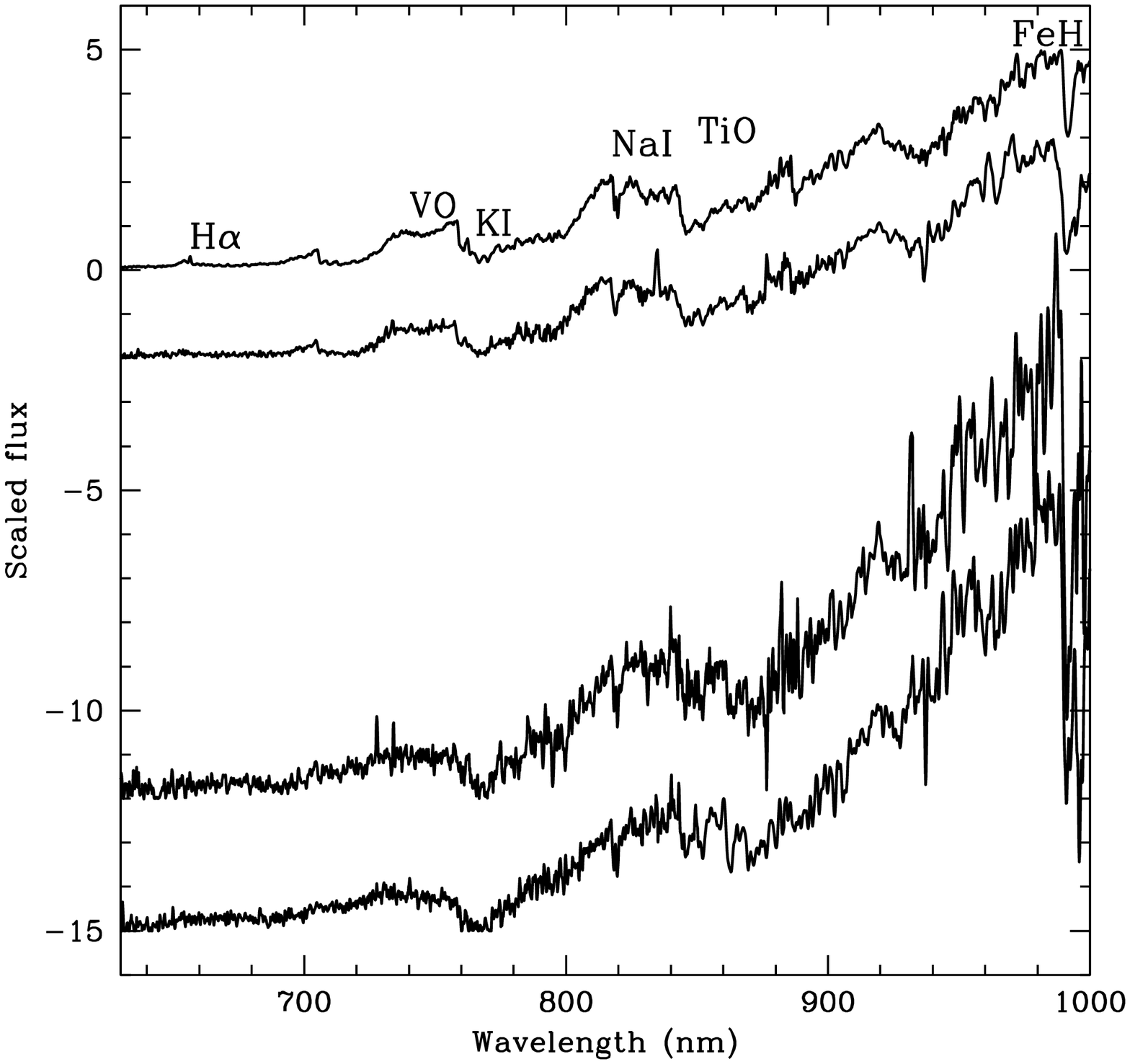}}
\caption{Full spectra for 5 UDs observed at SSO, including two of our objects 
with the latest spectral types. From top to bottom we show the spectra of 
VB10 (dM8), DENIS J1633131$-$755322 (dM9.5), 
DENIS J1206501$-$393725 (dL2), and DENIS J0014554$-$484417 (dL2.5). 
Some of the main spectral features discussed in this paper are labeled.} 
\end{figure}

\begin{figure}
\vskip 1in
\hskip -0.25in
\centerline{\includegraphics[width=5in,angle=0]{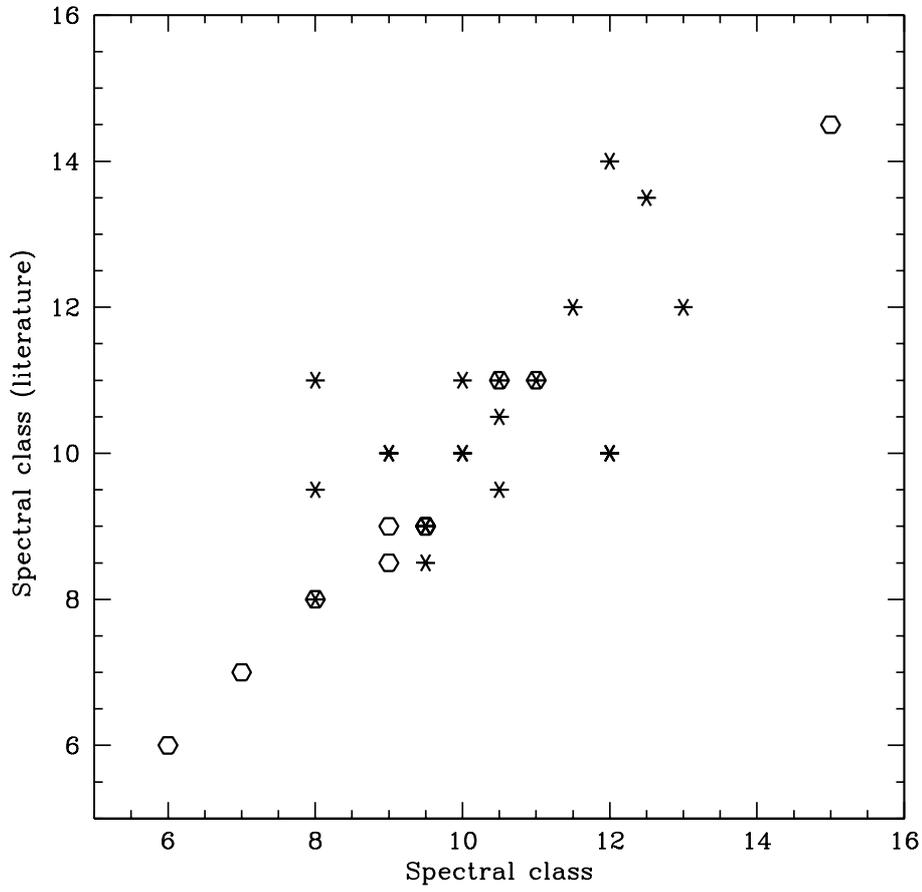}}
\caption{Comparison between our spectral types and those 
in the literature. Open symbols denote the objects from the literature observed 
by us as spectral type references. Six pointed skeletal symbols denote DENIS 
UD candidates that have published spectral classification.  
Generally, there is a good agreement within the standard uncertainty 
associated with spectral type determination ($\pm$0.5 subclass). }
\end{figure}

\begin{figure}
\vskip 1in
\hskip -0.25in
\centerline{\includegraphics[width=5in,angle=0]{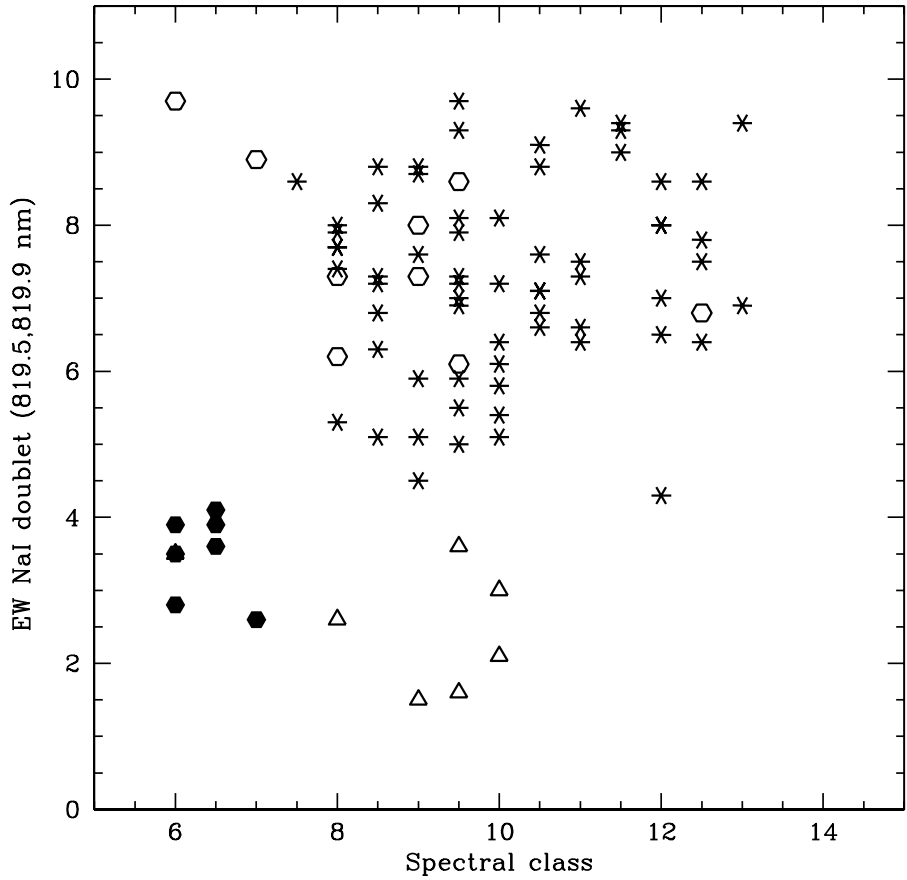}}
\caption{Equivalent widths of the NaI doublet (given in Table 4) versus spectral 
type (listed in Table 3) for our 65 high-gravity program field objects 
(six pointed skeletal symbol), our 6 low-gravity program objects (open triangles),  
our 7 Upper Sco candidates (solid hexagons) and our 12 
reference objects (open hexagons). }
\end{figure}

\begin{figure}
\vskip 1in
\hskip -0.25in
\centerline{\includegraphics[width=5in,angle=0]{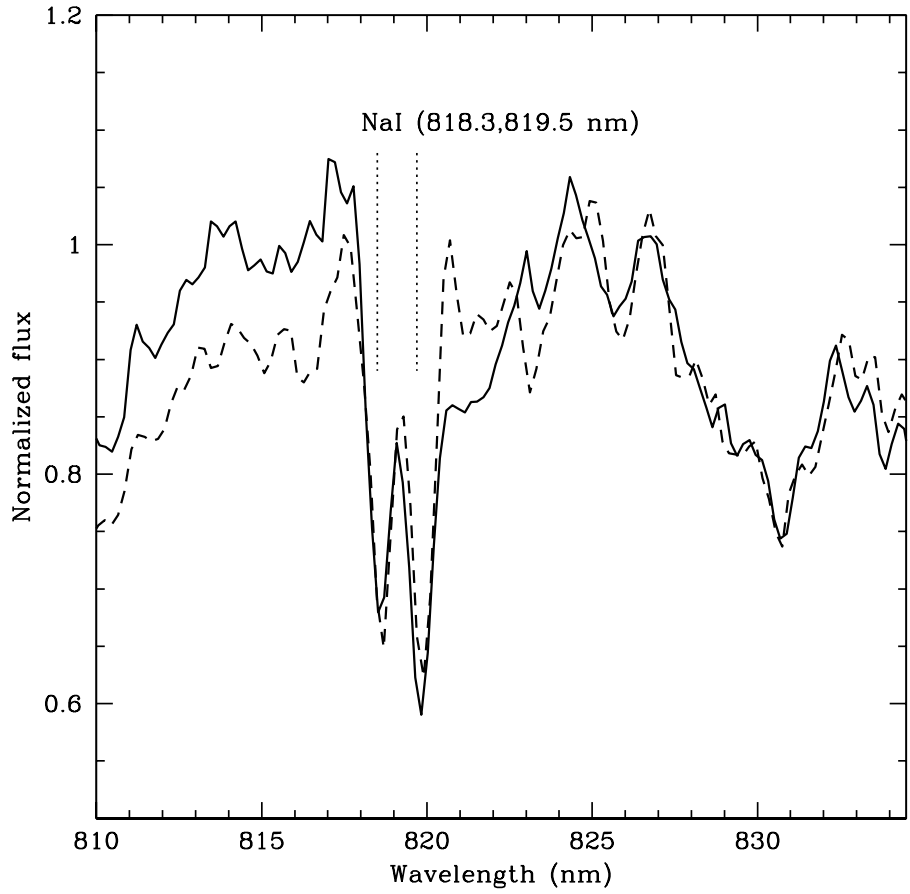}}
\caption{Comparison of our two spectra of  VB10 in the region around the NaI subordinate doublet.  
The solid line is the SSO spectrum (spectral resolution 8.5 \AA ) and the dashed line 
is the CTIO spectrum (spectral resolution 7.5 \AA ). The NaI doublet 
appears to be wider in the SSO spectrum because of blending effects with 
other absorption features.  }
\end{figure}

\begin{figure}
\vskip 1in
\hskip -0.25in
\centerline{\includegraphics[width=5in,angle=0]{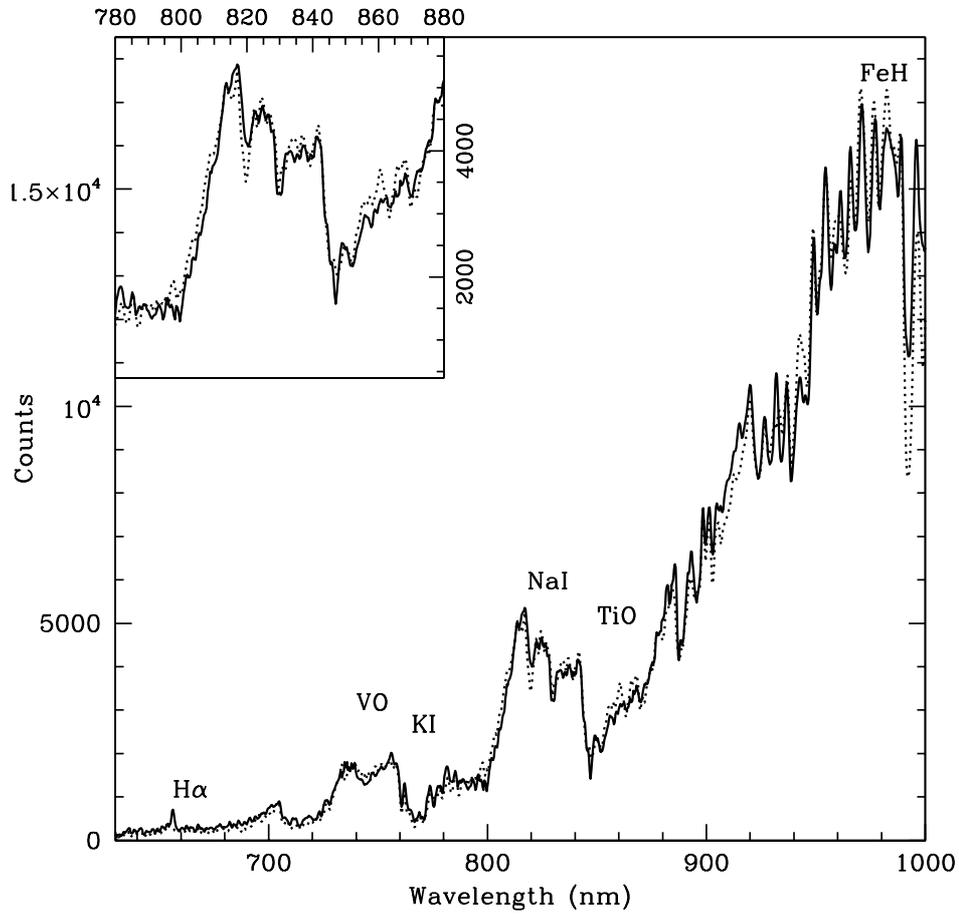}}
\caption{Comparison of the SSO spectrum of DENIS J0006579$-$643654 (solid line)  
with the SSO spectrum of DENIS J2150133$-$661036 (dotted line) which was observed with the same 
instrumental setup and has the same spectral class (dL0). A zoom of the NaI spectral region 
is displayed in the upper left corner.}
\end{figure}

\begin{figure}
\vskip 1in
\hskip -0.25in
\centerline{\includegraphics[width=5in,angle=0]{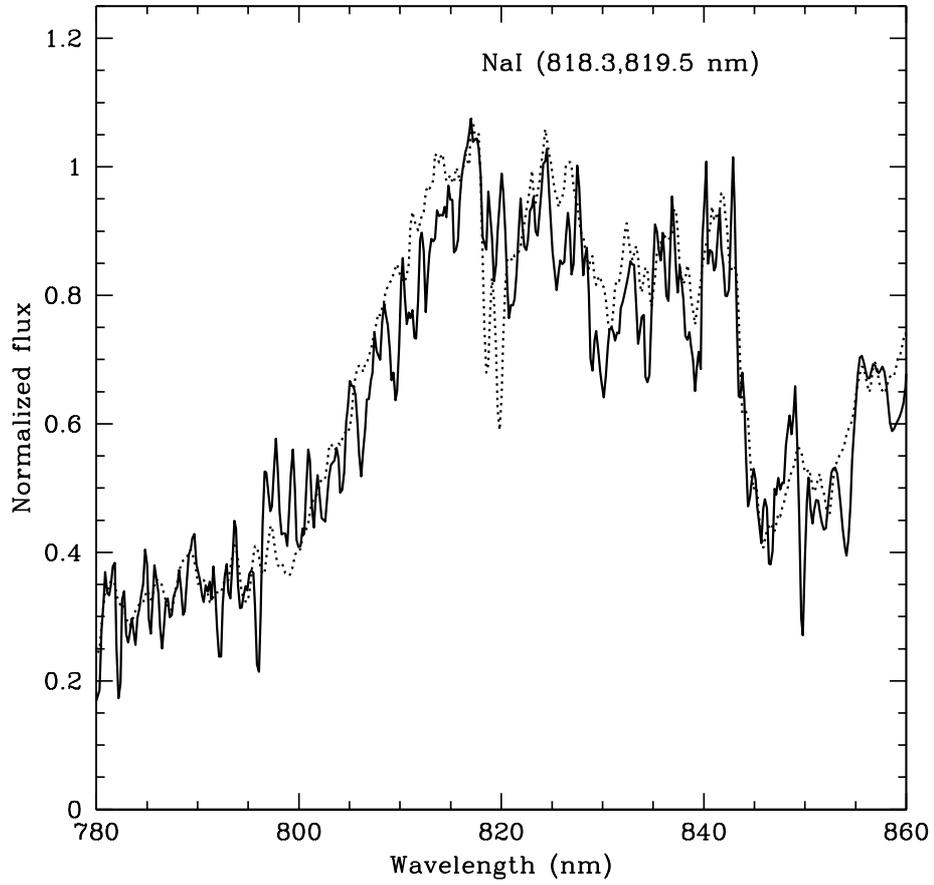}}
\caption{Comparison of the SSO spectrum of DENIS J1901391$-$370017 (solid) with the spectrum of VB10 
(dotted) obtained 
with the same instrumental setup. Both objects have the same spectral class (M8) but display 
different NaI subordinate doublet absorption.  }
\end{figure}

\begin{figure}
\vskip 1in
\hskip -0.25in
\centerline{\includegraphics[width=5in,angle=0]{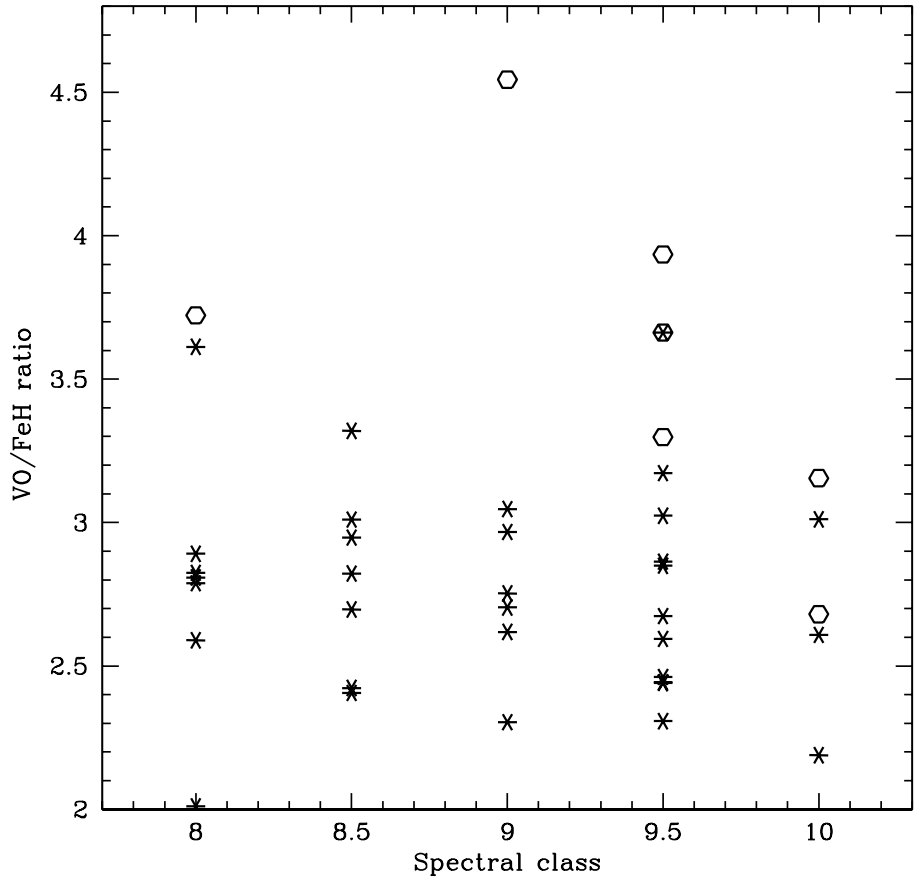}}
\caption{Ratio of the VO versus FeH molecular band indices with respect to spectral subclass. 
The low-gravity objects are denoted with open hexagons and tend to display higher values than 
the rest of the sample.  }
\end{figure}

\begin{figure}
\vskip 1in
\hskip -0.25in
\centerline{\includegraphics[width=5in,angle=0]{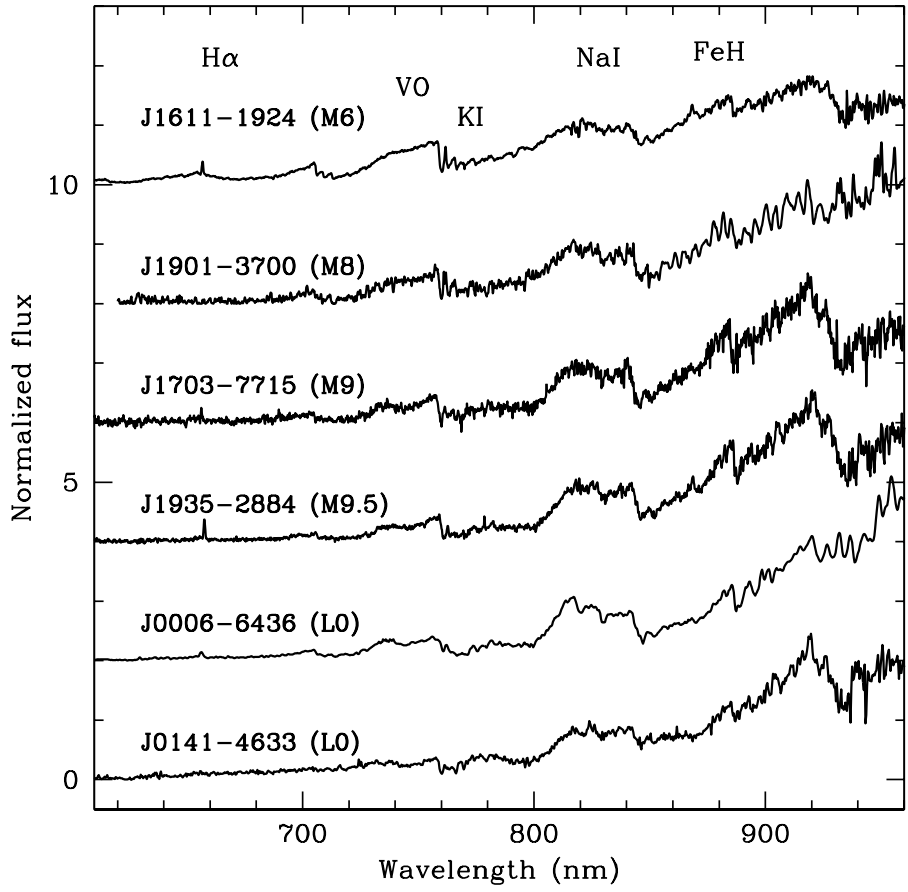}}
\caption{Full spectra of five of our low-gravity UDs and one of our Upper Sco BDs.   }
\end{figure}

\begin{figure}
\vskip 1in
\hskip -0.25in
\centerline{\includegraphics[width=5in,angle=0]{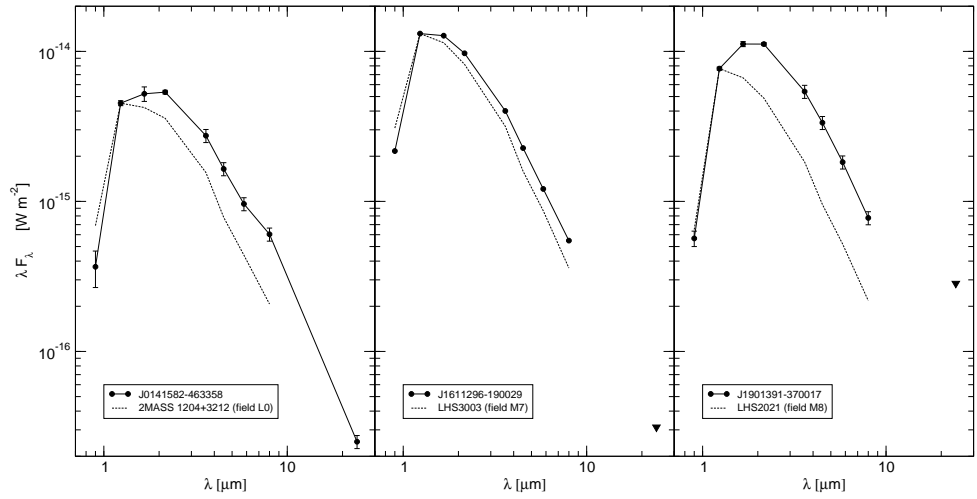}}
\caption{Spectral energy distribution of our three young BDs that have been observed with Spitzer. 
Shown for comparison (dotted lines) are the spectral energy distributions of three normal dwarfs of similar 
spectral class. }
\end{figure}

\begin{figure} 
\vskip 1in 
\hskip -0.25in
\centerline{\includegraphics[width=5in,angle=0]{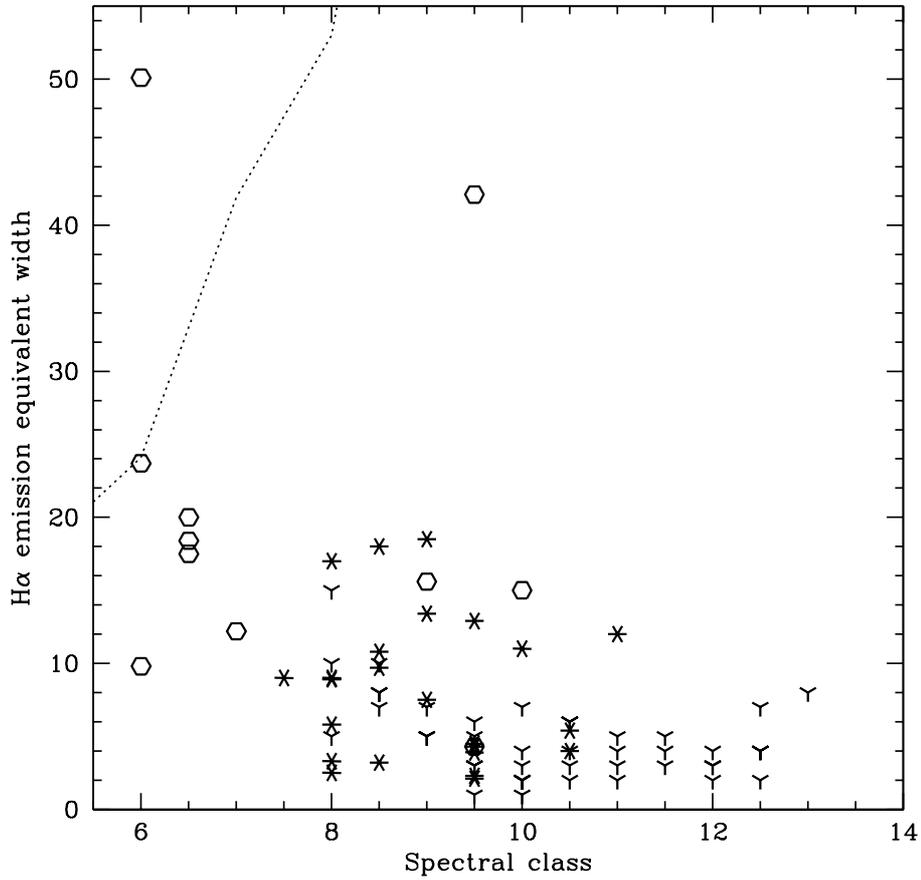}} 
\caption{H$\alpha$ equivalent width versus spectral class in our sample. 
Lower-gravity objects are denoted with open hexagons. Higher-gravity objects with  H$\alpha$ 
detection are denoted with 6-pointed skeleton symbols. Objects for which  H$\alpha$  was not 
detected in our spectra are shown as 3-pointed skeletons (upper limits). 
The dotted line is the boundary between accretors and non-accretors defined by 
Barrado y Navascu\'es \&  Mart\'{\i}n (2003).  }
\end{figure}

\clearpage

\begin{table*}
    \caption{Photometric data for 78 DENIS ultracool dwarf candidates}
    \label{photo}
  $$
\begin{tabular}{lllllllll}
   \hline 
   \hline
   \noalign{\smallskip}
DENIS name  & RA (J2000) &  Dec (J2000) &  $I$ &  $I-J$ &  $J-K$ &  errI &  errJ &  errK  \\
(1)         &  (2)       &  (3)         &  (4) &  (5)   &  (6)   &  (7)  &  (8)  &  (9)   \\
   \hline
    \noalign{\smallskip}
J0000286$-$124514 & 00 00 28.7 & $-$12 45 14 &  16.16 & 3.09 & 1.05 & 0.08 & 0.11 & 0.11 \\
J0006579$-$643654 & 00 06 57.9 & $-$64 36 54 &  16.71 & 3.28 & 1.32 & 0.12 & 0.09 & 0.11 \\
J0014554$-$484417 & 00 14 55.4 & $-$48 44 17 &  17.54 & 3.56 & 1.40 & 0.13 & 0.08 & 0.13 \\
J0028554$-$192716 & 00 28 55.4 & $-$19 27 16 &  17.61 & 3.66 & 1.18 & 0.18 & 0.18 & 0.17 \\
J0031192$-$384035 & 00 31 19.3 & $-$38 40 35 &  17.62 & 3.52 & 1.24 & 0.14 & 0.10 & 0.15 \\
J0050244$-$153818 & 00 50 24.4 & $-$15 38 18 &  16.86 & 3.20 & 1.10 & 0.10 & 0.10 & 0.11 \\
J0053189$-$363110 & 00 53 19.0 & $-$36 31 10 &  18.10 & 3.89 & 1.30 & 0.19 & 0.12 & 0.16 \\
J0055005$-$545026 & 00 55 00.5 & $-$54 50 26 &  17.12 & 3.38 & 1.00 & 0.11 & 0.20 & 0.15 \\
J0116529$-$645557 & 01 16 52.9 & $-$64 55 57 &  17.90 & 3.46 & 1.28 & 0.16 & 0.10 & 0.17 \\
J0128266$-$554534 & 01 28 26.6 & $-$55 45 34 &  17.07 & 3.26 & 1.47 & 0.16 & 0.12 & 0.13 \\
J0141582$-$463358 & 01 41 58.2 & $-$46 33 58 &  18.37 & 3.61 & 1.84 & 0.25 & 0.12 & 0.17 \\
J0147327$-$495448 & 01 47 32.8 & $-$49 54 48 &  16.05 & 3.14 & 0.97 & 0.07 & 0.09 & 0.07 \\
J0206566$-$073519 & 02 06 56.7 & $-$07 35 20 &  17.92 & 3.58 & 1.35 & 0.17 & 0.11 & 0.15 \\
J0213371$-$134322 & 02 13 37.1 & $-$13 43 22 &  17.64 & 3.36 & 1.03 & 0.17 & 0.16 & 0.18 \\
J0224120$-$763320 & 02 24 12.0 & $-$76 33 20 &  18.07 & 2.71 & 2.05 & 0.19 & 0.18 & 0.19 \\
J0227102$-$162446 & 02 27 10.2 & $-$16 24 46 &  17.04 & 3.37 & 1.49 & 0.11 & 0.12 & 0.18 \\
J0230450$-$095305 & 02 30 45.0 & $-$09 53 05 &  18.24 & 3.56 & 1.69 & 0.21 & 0.18 & 0.15 \\
J0240121$-$530552 & 02 40 12.1 & $-$53 05 52 &  18.18 & 3.83 & 1.36 & 0.22 & 0.10 & 0.16 \\
J0301488$-$590302 & 03 01 48.8 & $-$59 03 02 &  16.80 & 3.37 & 1.11 & 0.10 & 0.08 & 0.09 \\
J0314352$-$462340 & 03 14 35.2 & $-$46 23 41 &  17.99 & 3.13 & 1.15 & 0.20 & 0.15 & 0.20 \\
J0325293$-$431229 & 03 25 29.4 & $-$43 12 30 &  17.48 & 3.32 & 1.21 & 0.15 & 0.10 & 0.16 \\
J0357290$-$441730 & 03 57 29.0 & $-$44 17 31 &  17.91 & 3.39 & 1.73 & 0.19 &	0.16 & 0.17 \\
J0427270$-$112713 & 04 27 27.1 & $-$11 27 14 &  16.67 & 3.14 & 0.99 & 0.13 & 0.08 & 0.12 \\
J0428510$-$225322 & 04 28 51.0 & $-$22 53 22 &  16.80 & 3.35 & 1.48 & 0.10 & 0.08 & 0.11 \\
J0436360$-$295947 & 04 36 36.0 & $-$29 59 47 &  18.14 & 3.33 & 1.31 & 0.23 &	0.14 & 0.21 \\
J0443376$+$000205 & 04 43 37.6 & $+$00 02 05 &  15.88 & 3.35 & 1.42 & 0.05 & 0.10 & 0.11 \\
J0529572$-$200300 & 05 29 57.2 & $-$20 03 00 &  17.84 & 3.39 & 1.06 & 0.25 &	0.15 & 0.20 \\
J0608528$-$275358 & 06 08 52.8 & $-$27 53 58 &  17.09 & 3.41 & 1.47 & 0.10 &	0.09 & 0.11 \\
J0610008$-$472741 & 06 10 00.9 & $-$47 27 41 &  17.46 & 3.00 & 1.29 & 0.15 & 0.13 & 0.17 \\
J0620165$-$430009 & 06 20 16.5 & $-$43 00 09 &  17.78 & 2.82 &      & 0.18 & 0.13 &     \\ 
J0719317$-$505141 & 07 19 31.8 & $-$50 51 41 &  17.44 & 3.44 & 1.09 & 0.11 & 0.09 & 0.14 \\
J0921141$-$210445 & 09 21 14.1 & $-$21 04 45 &  16.50 & 3.65 & 1.02 & 0.09 & 0.08 & 0.10 \\
J0953213$-$101420 & 09 53 21.3 & $-$10 14 20 &  16.82 & 3.30 & 1.41 & 0.10 & 0.08 & 0.11 \\
J1004283$-$114648 & 10 04 28.3 & $-$11 46 48 &  18.02 & 3.17 &      & 0.20 & 0.15 &      \\
J1004403$-$131818 & 10 04 40.3 & $-$13 18 19 &  17.80 & 3.14 & 1.49 & 0.18 & 0.24 & 0.15 \\
J1019245$-$270717 & 10 19 24.6 & $-$27 07 17 &  16.90 & 3.33 & 1.08 & 0.10 & 0.08 & 0.15 \\
J1115297$-$242934 & 11 15 29.7 & $-$24 29 35 &  16.50 & 3.12 & 0.93 & 0.08 & 0.07 & 0.15 \\ 
J1206501$-$393725 & 12 06 50.1 & $-$39 37 26 &  17.67 & 3.36 & 1.19 & 0.16 & 0.10 & 0.16 \\ 
J1216121$-$125731 & 12 16 12.1 & $-$12 57 31 &  18.30 & 3.18 &      & 0.23 & 0.26 &      \\ 
J1232183$-$095149 & 12 32 18.3 & $-$09 51 50 &  16.97 & 3.20 & 1.34 & 0.10 & 0.12 & 0.12 \\ 
J1234018$-$112407 & 12 34 01.9 & $-$11 24 07 &  18.22 & 3.59 & 1.23 & 0.19 & 0.13 & 0.20 \\ 
J1256569$+$014616 & 12 56 56.9 & $+$01 46 17 &  18.20 & 3.77 & 1.49 & 0.17 & 0.09 & 0.13 \\ 
J1359551$-$403456 & 13 59 55.1 & $-$40 34 56 &  16.98 & 3.20 & 1.16 & 0.10 & 0.10 & 0.13 \\ 
J1411051$-$791536 & 14 11 05.2 & $-$79 15 36 &  16.23 & 3.10 & 1.06 & 0.07 & 0.08 & 0.10 \\ 
J1555256$-$181748 & 15 55 25.6 & $-$18 17 48 &  14.75 & 2.35 & 1.09 & 0.05 & 0.10 & 0.10 \\ 
J1600256$-$192750 & 16 00 25.6 & $-$19 27 50 &  14.68 & 2.42 & 1.11 & 0.05 & 0.10 & 0.10 \\ 
J1602043$-$205043 & 16 02 04.3 & $-$20 50 43 &  15.16 & 2.40 & 0.94 & 0.05 & 0.10 & 0.10 \\ 
J1602553$-$192243 & 16 02 55.3 & $-$19 22 43 &  15.07 & 2.40 & 0.98 & 0.05 & 0.10 & 0.10  \\ 
J1611014$-$192449 & 16 11 01.4 & $-$19 24 49 &  15.69 & 2.41 & 0.91 & 0.05 & 0.10 & 0.10  \\ 
J1611124$-$192737 & 16 11 12.4 & $-$19 27 37 &  15.20 & 2.41 & 1.11 & 0.05 & 0.10 & 0.10   \\
J1611296$-$190029 & 16 11 29.6 & $-$19 00 29 &  16.46 & 2.80 & 1.19 & 0.05 & 0.10 & 0.10 \\ 
J1622326$-$120719 & 16 22 32.7 & $-$12 07 19 &  16.56 & 3.20 & 0.89 & 0.07 & 0.08 & 0.13 \\ 
J1633131$-$755322 & 16 33 13.1 & $-$75 53 23 &  16.20 & 3.10 & 1.10 & 0.06 & 0.07 & 0.10 \\ 
J1703356$-$771520 & 17 03 35.6 & $-$77 15 20 &  18.25 & 3.09 &      & 0.15 & 0.10 &  \\ 
J1707252$-$013809 & 17 07 25.2 & $-$01 38 09 &  17.81 & 3.55 & 1.40 & 0.15 & 0.12 & 0.14 \\ 
J1716352$-$031542 & 17 16 35.2 & $-$03 15 42 &  14.46 & 3.43 & 1.71 & 0.05 & 0.10 & 0.09 \\ 
J1753452$-$655955 & 17 53 45.2 & $-$65 59 55 &  17.80 & 3.59 & 1.79 & 0.14 & 0.10 & 0.12 \\ 
    \noalign{\smallskip}
    \hline 
   \end{tabular}
 $$
  \begin{list}{}{}
  \item[]
  \end{list}
\end{table*} 

\clearpage

\setcounter{table}{0}
\begin{table*}
    \caption{continued.}
    \label{photo}
  $$
\begin{tabular}{lllllllll}
   \hline 
   \hline
   \noalign{\smallskip}
DENIS name  & RA (J2000) &  Dec (J2000) &  $I$ &  $I-J$ &  $J-K$ &  errI &  errJ &  errK  \\
(1)         &  (2)       &  (3)         &  (4) &  (5)   &  (6)   &  (7)  &  (8)  &  (9)   \\
   \hline
    \noalign{\smallskip}
J1901391$-$370017 & 19 01 39.1 & $-$37 00 17 &  17.94 & 3.71 & 1.94 & 0.16 & 0.11 & 0.10 \\ 
J1907440$-$282420 & 19 07 44.0 & $-$28 24 20 &  17.95 & 3.60 & 0.97 & 0.16 & 0.12 & 0.18 \\ 
J1926005$-$650006 & 19 26 00.5 & $-$65 00 06 &  17.90 & 3.35 & 1.51 & 0.15 & 0.12 & 0.18 \\ 
J1934511$-$184134 & 19 34 51.2 & $-$18 41 35 &  17.71 & 3.43 & 1.15 & 0.14 & 0.11 & 0.16 \\ 
J1935560$-$284634 & 19 35 56.0 & $-$28 46 34 &  17.21 & 3.30 & 1.23 & 0.14 & 0.11 & 0.16 \\ 
J1956460$-$774717 & 19 56 46.0 & $-$77 47 17 &  17.46 & 3.28 & 1.14 & 0.13 & 0.11 & 0.18 \\ 
J2013108$-$124244 & 20 13 10.8 & $-$12 42 45 &  18.07 & 3.55 & 1.21 & 0.17 & 0.15 & 0.17 \\ 
J2030412$-$363509 & 20 30 41.2 & $-$36 35 09 &  17.50 & 3.19 & 1.21 & 0.16 & 0.09 & 0.12 \\ 
J2045024$-$633206 & 20 45 02.4 & $-$63 32 06 &  16.05 & 3.40 & 1.45 & 0.13 & 0.12 & 0.15 \\ 
J2126340$-$314322 & 21 26 34.0 & $-$31 43 22 &  16.26 & 3.06 & 0.91 & 0.07 & 0.13 & 0.16 \\ 
J2139136$-$352950 & 21 39 13.6 & $-$35 29 51 &  17.94 & 3.47 & 1.11 & 0.17 & 0.11 & 0.20 \\ 
J2143510$-$833712 & 21 43 51.0 & $-$83 37 12 &  16.50 & 3.30 & 0.98 & 0.10 & 0.06 & 0.11 \\ 
J2150133$-$661036 & 21 50 13.3 & $-$66 10 37 &  17.23 & 3.55 & 1.13 & 0.14 & 0.08 & 0.12 \\ 
J2150149$-$752035 & 21 50 15.0 & $-$75 20 36 &  17.45 & 3.51 & 1.39 & 0.16 & 0.09 & 0.12 \\ 
J2243169$-$593219 & 22 43 17.0 & $-$59 32 20 &  17.40 & 3.32 & 1.11 & 0.15 & 0.10 & 0.14 \\ 
J2308113$-$272200 & 23 08 11.3 & $-$27 22 01 &  18.11 & 3.53 & 1.40 & 0.20 & 0.17 & 0.17 \\ 
J2322468$-$313323 & 23 22 46.8 & $-$31 33 23 &  16.84 & 3.29 & 1.26 & 0.15 & 0.13 & 0.17 \\ 
J2329343$-$540858 & 23 29 34.3 & $-$54 08 58 &  18.37 & 3.41 & 1.74 & 0.20 & 0.12 & 0.21 \\ 
J2330226$-$034717 & 23 30 22.6 & $-$03 47 17 &  17.74 & 3.33 & 1.28 & 0.10 & 0.15 & 0.20 \\ 
J2345390$+$005514 & 23 45 39.0 & $+$00 55 14 &  16.90 & 3.18 & 1.28 & 0.12 & 0.13 & 0.13 \\ 
J2354599$-$185221 & 23 54 59.9 & $-$18 52 21 &  17.43 & 3.21 & 1.39 & 0.11 & 0.09 & 0.13 \\ 
    \noalign{\smallskip}
    \hline 
   \end{tabular}
 $$
  \begin{list}{}{}
  \item[]
  \end{list}
\end{table*} 

\clearpage

\begin{table*}
    \caption{Spectroscopic observing log}
    \label{obs}
  $$
\begin{tabular}{llllll}
   \hline 
   \hline
   \noalign{\smallskip}
Name  &  Telescope &  Date &  Texp  & Disp. & Res.             \\
(1)   &  (2)       &  (3)  &  (4)   &  (5)  &  (6)             \\               
   \hline
    \noalign{\smallskip}
DENIS J0000286$-$124514 & SSO 2.3m & 23 June 2006 & 900 & 3.70 & 12.5  \\  
DENIS J0006579$-$643654 & SSO 2.3m & 23 June 2006 & 900 & 3.70 & 12.5 \\  
DENIS J0014554$-$484417 & SSO 2.3m & 27 June 2006 & 1200 & 1.87 & 8.5 \\ 
DENIS J0028554$-$192716 & SSO 2.3m & 27 June 2006 & 1200 & 1.87 & 8.5 \\  
DENIS J0031192$-$384035 & SSO 2.3m & 27 June 2006 & 1200 & 1.87 & 8.5 \\   
                         & NTT      & 30 Dec 2000  & 2700 & 2.73 & 10.5 \\  
DENIS J0050244$-$153818 & SSO 2.3m & 27 June 2006 & 1200 & 1.87 & 8.5 \\  
                         & NTT      & 30 Dec 2000  & 1800 & 2.73 & 10.5 \\ 
DENIS J0053189$-$363110 & SSO 2.3m & 28 June 2006 & 1200 & 1.87 & 8.5 \\ 
DENIS J0055005$-$545026 & SSO 2.3m & 27 June 2006 & 1200 & 1.87 & 8.5 \\ 
DENIS J0116529$-$645557 & SSO 2.3m & 27 June 2006 & 1200 & 1.87 & 8.5 \\ 
DENIS J0128266$-$554534 & SSO 2.3m & 24 June 2006 & 1200 & 1.87 & 8.5 \\ 
DENIS J0141582$-$463358 & NTT      & 29 Nov 2003  & 2400 & 3.62 & 8.5 \\ 
DENIS J0147327$-$495448 & SSO 2.3m & 24 June 2006 &  600 & 1.87 & 8.5 \\ 
DENIS J0206566$-$073519 & SSO 2.3m & 27 June 2006 & 1200 & 1.87 & 8.5 \\ 
DENIS J0213371$-$134322 & SSO 2.3m & 27 June 2006 & 1200 & 1.87 & 8.5 \\ 
DENIS J0224120$-$763320 & NTT      & 30 Dec 2000  & 2700 & 2.73 & 10.5 \\ 
DENIS J0227102$-$162446 & SSO 2.3m & 27 June 2006 &  900 & 1.87 & 8.5 \\ 
DENIS J0230450$-$095305 & SSO 2.3m & 28 June 2006 & 1200 & 1.87 & 8.5 \\ 
DENIS J0240121$-$530552 & SSO 2.3m & 28 June 2006 & 1200 & 1.87 & 8.5 \\ 
DENIS J0301488$-$590302 & SSO 2.3m & 27 June 2006 & 1200 & 1.87 & 8.5 \\ 
DENIS J0314352$-$462341 & VLT      & 12 Aug 2002  & 500  & 1.31 & 3.3 \\ 
DENIS J0325293$-$431229 & SSO 2.3m & 24 June 2006 & 1700 & 1.87 & 8.5 \\ 
DENIS J0357290$-$441731 & VLT      & 12 Aug 2002  & 500 & 1.31  & 3.3 \\ 
DENIS J0427271$-$112713 & WHT      & 5 Dec 2006   & 1200 & 1.63 & 6.5 \\ 
DENIS J0428510$-$225322 & NTT      & 29 Nov 2003  & 2000 & 3.62 & 8.5 \\ 
DENIS J0436360$-$295947 & NTT      & 30 Dec 2000  & 2700 & 2.73 & 10.5 \\ 
DENIS J0443373$+$000205 & NTT      & 30 Dec 2000  &  900 & 2.73 & 10.5 \\ 
DENIS J0529572$-$200300 & NOT      & 9 March 2000 & 4800 & 3.10 & 20.0 \\ 
                         & WHT      & 29 Sept 2000 & 900 & 2.90 & 6.5 \\
DENIS J0608528$-$275358 & WHT      & 28 Sept 2000 & 1200 & 2.90 & 6.5 \\ 
DENIS J0610008$-$472741   & NTT      & 30 Dec 2000  & 1800 & 2.73 & 10.5 \\ 
DENIS J0620165$-$430010   & NTT      & 30 Dec 2000  & 2700 & 2.73 & 10.5 \\ 
DENIS J0719317$-$505141 & SSO 2.3m & 24 June 2006 & 900 & 3.70 & 12.5 \\ 
DENIS J0921141$-$210445 & SSO 2.3m & 24 June 2006 & 600 & 3.70 & 12.5 \\ 
DENIS J0953213$-$101420 & NTT      & 30 Nov 2003  & 1500 & 3.62 & 8.5 \\ 
DENIS J1004283$-$114648  & NOT      &  9 March 2000  & 4800 & 3.10 & 20.0 \\ 
                         & VLT      & 31 Dec 2002  & 1600 & 0.73 & 2.1 \\ 
DENIS J1004403$-$131818 & NOT      & 10 March 2000  & 2400 & 3.10 & 20.0 \\ 
DENIS J1019245$-$270717 & SSO 2.3m & 24 June 2006 & 600 & 3.70 & 12.5 \\ 
DENIS J1115297$-$242934 & SSO 2.3m & 22 June 2006 & 900 & 3.70  & 12.5 \\ 
DENIS J1206501$-$393725 & SSO 2.3m & 25 June 2006 & 1200 & 1.87 & 8.5 \\ 
DENIS J1216121$-$125731 & NOT      & 10 March 2000 & 5400 & 3.10 & 20.0 \\
DENIS J1232183$-$095149 & NOT      & 10 March 2000 & 2260 & 3.10 & 20.0 \\ 
DENIS J1234018$-$112407 & SSO 2.3m & 29 June 2006 & 1200 & 1.87 & 8.5 \\ 
DENIS J1256569$+$014616 & SSO 2.3m & 29 June 2006 & 1200 & 1.87 & 8.5 \\ 
DENIS J1359551$-$403456 & SSO 2.3m & 24 June 2006 & 1200 & 1.87 & 8.5 \\ 
DENIS J1411051$-$791536 & SSO 2.3m & 22 June 2006 & 900 & 3.70  & 12.5 \\ 
DENIS J1555256$-$181748 & CTIO 4m  & 17 July  2007 & 1200 & 2.01 & 7.5 \\ 
DENIS J1600256$-$192750 & CTIO 4m  & 17 July  2007 & 1600 & 2.01 & 7.5 \\ 
DENIS J1602043$-$205043 & CTIO 4m  & 17 July  2007 & 1600 & 2.01 & 7.5 \\ 
DENIS J1602553$-$192243 & CTIO 4m  & 17 July  2007 & 1200 & 2.01 & 7.5 \\ 
DENIS J1611014$-$192449 & CTIO 4m  & 17 July  2007 & 1600 & 2.01 & 7.5 \\ 
DENIS J1611124$-$192737 & CTIO 4m  & 17 July  2007 & 1600 & 2.01 & 7.5 \\ 
DENIS J1611296$-$190029 & CTIO 4m  & 17 July  2007 & 2000 & 2.01 & 7.5 \\ 
DENIS J1622326$-$120719 & SSO 2.3m & 24 June 2006  & 900  & 3.70  & 12.5 \\ 
DENIS J1633131$-$755322 & SSO 2.3m & 22 June 2006  & 1200 & 3.70 & 12.5 \\ 
DENIS J1703356$-$771520 & CTIO 4m  & 17 July  2007 & 1600 & 2.01 & 7.5 \\ 
DENIS J1707252$-$013809 & SSO 2.3m & 26 June 2006 & 1200 & 1.87 & 8.5 \\ 
DENIS J1716352$-$031542 & SSO 2.3m & 22 June 2006 &  600 & 3.70 & 12.5 \\ 
DENIS J1753452$-$655955 & SSO 2.3m & 26 June 2006 & 1200 & 1.87 & 8.5 \\ 
    \noalign{\smallskip}
    \hline 
   \end{tabular}
 $$
  \begin{list}{}{}
  \item[]
  \end{list}
\end{table*} 

\clearpage

\setcounter{table}{1}
\begin{table*}
    \caption{continued.}
    \label{obs}
  $$
\begin{tabular}{llllll}
   \hline 
   \hline
   \noalign{\smallskip}
Name  &  Telescope &  Date &  Texp  & Disp. & Res.             \\
(1)   &  (2)       &  (3)  &  (4)   &  (5)  &  (6)             \\               
   \hline
    \noalign{\smallskip}
DENIS J1901391$-$370017 & SSO 2.3m & 27 June 2006 & 1800 & 1.87 & 8.5 \\ 
DENIS J1907440$-$282420 & SSO 2.3m & 27 June 2006 & 1800 & 1.87 & 8.5 \\ 
DENIS J1926005$-$650006 & CTIO 4m  & 17 July  2007 & 1600 & 2.01 & 7.5 \\ 
DENIS J1934511$-$184134 & SSO 2.3m & 25 June 2006 & 1200 & 1.87 & 8.5 \\ 
DENIS J1935560$-$284634 & CTIO 4m  & 17 July  2007 & 1600 & 2.01 &  7.5 \\ 
DENIS J1956460$-$774717 & CTIO 4m  & 17 July  2007 & 1600 & 2.01 &  7.5 \\ 
DENIS J2013108$-$124244 & SSO 2.3m & 28 June 2006 & 1800 & 1.87 & 8.5 \\ 
DENIS J2030412$-$363509 & CTIO 4m  & 17 July  2007 & 1600 & 2.01 & 7.5 \\ 
DENIS J2045024$-$633206 & NTT      & 29 Nov 2003  & 600  & 3.62 & 8.5 \\ 
DENIS J2126340$-$314322 & SSO 2.3m & 22 June 2006 & 1200 & 3.70 & 12.5 \\ 
DENIS J2139136$-$352950 & SSO 2.3m & 25 June 2006 & 1200 & 1.87 & 8.5 \\ 
DENIS J2143510$-$833712 & SSO 2.3m & 25 June 2006 & 1200 & 3.70 & 12.5 \\ 
DENIS J2150133$-$661036 & SSO 2.3m & 23 June 2006 & 1200 & 3.70 & 12.5 \\ 
DENIS J2150149$-$752035 & SSO 2.3m & 23 June 2006 & 1200 & 3.70 & 12.5 \\ 
DENIS J2243169$-$593219 & SSO 2.3m & 23 June 2006 & 1200 & 3.70 & 12.5 \\ 
DENIS J2308113$-$272200 & SSO 2.3m & 28 June 2006 & 1200 & 1.87 & 8.5 \\ 
DENIS J2322468$-$313323 & WHT      & 29 Sept 2000 & 1800 & 2.90 & 6.5 \\ 
DENIS J2329343$-$540854 & VLT      & 18 July 2002 & 500 & 1.31 & 3.3 \\ 
DENIS J2330226$-$034717 & WHT      & 29 Sept 2000 & 1800 & 2.90 & 6.5 \\ 
DENIS J2345390$+$005514 & SSO 2.3m & 25 June 2006 & 1200 & 1.87 & 8.5 \\ 
DENIS J2354599$-$185221 & WHT      & 29 Sept 2000 & 1800 & 2.90 & 6.5 \\ 
GJ 406                    & WHT      & 5 Dec 2006   &  300 & 1.63 & 6.5 \\ 
LHS 2397a                 & SSO 2.3m & 25 June 2006 & 400  & 1.87 & 8.5 \\ 
LHS 2924                  & SSO 2.3m & 26 June 2006 & 900  & 1.87 & 8.5 \\ 
LP 944-20                & WHT      & 28 Sept 2000 & 300  & 2.90 & 6.5 \\ 
VB 8                      & SSO 2.3m & 25 June 2006 & 360  & 1.87 & 8.5 \\ 
VB 10                     & SSO 2.3m & 25 June 2006 & 600  & 1.87 & 8.5 \\ 
                         & CTIO 4m  & 17 July  2007 & 300 & 2.01 & 7.5 \\ 
DENIS J1048147$-$395606 & NTT      & 30 Nov 2003  & 60   & 3.62 & 8.5 \\ 
DENIS J1228152$-$154733   & VLT      & 17 June 2002 & 500 & 1.31 & 3.3 \\ 
DENIS J1441373$-$094559 & NOT      & 10 March 2000 & 4800 & 3.10 & 20.0 \\ 
2MASS 003615+182112      & WHT      & 28 Sept 2000 & 300  & 2.90 & 6.5 \\ 
    \noalign{\smallskip}
    \hline 
   \end{tabular}
 $$
  \begin{list}{}{}
  \item[]
  \end{list}
\end{table*}

\clearpage

\begin{table*}
    \caption{PC3 index and spectral type for DENIS ultracool candidates}
    \label{obs}
  $$
\begin{tabular}{llll}
   \hline 
   \hline
   \noalign{\smallskip}
DENIS or name &  PC3 &  SpT  &  Notes                 \\
 (1)          &  (2) &  (3)  &  (4)                   \\
   \hline
    \noalign{\smallskip}
J0000286$-$124514 & 2.35 & dM9.5 & M8.5 (1)  \\ 
J0006579$-$643654 & 2.42 & dL0   &  low-gravity         \\ 
J0014554$-$484417 & 3.64 & dL2.5 &           \\ 
J0028554$-$192716 & 2.77 & dL0.5 &           \\ 
J0031192$-$384035 & 3.70\footnotemark[1] & dL2.5 &           \\ 
J0050244$-$153818 & 2.65\footnotemark[2] & dL0.5 & L1: (1)   \\ 
J0053189$-$363110 & 4.03 & dL2.5 & L3.5 (2)  \\ 
J0055005$-$545026 & 1.94 & dM8.5 &           \\ 
J0116529$-$645557 & 2.93 & dL1   &           \\ 
J0128266$-$554534 & 2.84 & dL1   & L1 (3)    \\ 
J0141582$-$463358 & 2.57 & L0    & L0 (4); low-gravity \\ 
J0147327$-$495448 & 1.80 & dM8   & M8+L2 (5) \\ 
J0206566$-$073519 & 1.98 & dM8.5 &           \\ 
J0213371$-$134322 & 2.10 & dM9   &           \\ 
J0224120$-$763320 & 2.55 & dL0   &           \\ 
J0227102$-$162446 & 2.44 & dL0   & L1  (6)  \\ 
J0230450$-$095305 & 2.40 & dL0   &           \\ 
J0240121$-$530552 & 2.22 & dM9.5 &           \\ 
J0301488$-$590302 & 2.11 & dM9   & L0 (7)    \\ 
J0314352$-$462341 & 3.58 & dL2   & L0 (7)    \\ 
J0325293$-$431229 & 2.05 & dM8.5 &           \\ 
J0357290$-$441731 & 3.33 & dL2   & L0 (2); M9+L1.5 (8) \\ 
J0427271$-$112713 & 1.64 & dM7   &           \\ 
J0428510$-$225322 & 2.78 & dL0.5 & L0.5 (9)  \\ 
J0436360$-$295947 & 1.86 & dM8   &           \\ 
J0443373$+$000205 & 2.25 & dM9.5 & M9 (10); low-gravity   \\ 
J0529572$-$200300 & 2.10\footnotemark[3] & dM9.0 &  \\ 
J0608528$-$275358 & 2.21 & dM9.5 &   low-gravity        \\ 
J0610008$-$472741   & 2.01 & dM8.5 &           \\ 
J0620165$-$430010   & 1.75 & dM8   &           \\ 
J0719317$-$505141 & 2.87 & dL1   &           \\ 
J0921141$-$210445 & 4.54 & dL3   & L2 (6)    \\ 
J0953213$-$101420 & 2.45 & dL0   & L0 (1)    \\ 
J1004283$-$114648 & 1.89 & dM8   & M9.5+L0.5 (8) \\
J1004403$-$131818 & 2.35 & dL0   &           \\ 
J1019245$-$270717 & 2.74 & dL0.5 & M9.5 (3)  \\ 
J1115297$-$242934 & 1.87 & dM8   &           \\ 
J1206501$-$393725 & 3.15 & dL2   &           \\ 
J1216121$-$125731 & 1.76 & dM8   & L1 (7)    \\ 
J1232209$-$095102 & 0.97 &  M2   & giant     \\ 
J1234018$-$112407 & 2.36 & dM9.5 &           \\ 
J1256569$+$014616 & 2.98 & dL1.5 & L2   (6)  \\ 
J1359551$-$403456 & 3.08 & dL2   &           \\ 
J1411051$-$791536 & 2.02 & dM8.5 &           \\ 
J1555256$-$181748 & 1.48 &  M6   & low-gravity   \\ 
J1600256$-$192750 & 1.52 &  M6.5   & low-gravity   \\ 
J1602043$-$205043 & 1.50 &  M6.5   & low-gravity   \\ 
J1602553$-$192243 & 1.51 &  M6.5   & low-gravity   \\ 
J1611014$-$192449 & 1.41 &  M6   & low-gravity   \\ 
J1611124$-$192737 & 1.45 &  M6   & low-gravity   \\ 
J1611296$-$190029 & 1.61 &  M7   & low-gravity   \\ 
J1622326$-$120719 & 2.24 & dM9.5 &           \\ 
J1633131$-$755322 & 2.28 & dM9.5 &           \\ 
J1703356$-$771520 & 2.16 &  M9   & low-gravity \\ 
J1707252$-$013809 & 2.82 & dL0.5 &           \\ 
J1716352$-$031542 & 1.26 & M5    & giant     \\ 
J1753452$-$655955 & 3.47 & dL2   & L4 (6)    \\ 
    \noalign{\smallskip}
    \hline 
   \end{tabular}
 $$
  \begin{list}{}{}
  \item[]
Column 1: DENIS name.
Column 2: PC3 index defined in M99.
Colum 3: Spectral type from PC3-SpT relation in M99. 
Column 4: Notes about specific targets found in the literature. 
References: (1)=Cruz et al. 2007; 
(2)=Kirkpatrick et al., in preparation (DwarfArchives.org); 
(3)=Kendall et al. 2007; 
(4)=Kirkpatrick et al. 2006; 
(5)=Reid et al. 2006; 
(6)=Schmidt et al. 2007;
(7)=Bouy et al. 2003; 
(8)=Mart\'{\i}n et al. 2006; 
(9)=Kendall et al. 2003;
(10)=Hawley et al. 2002; 
(11)=Delfosse et al. 2001;
(12)=Mart\'{\i}n et al. 1999; 
(13)=Kirkpatrick et al. 2000  
\thanks{\footnotemark[1] Average value of two independent measurements; 
PC3=3.75 (SSO) and PC3=3.65 (NTT)}
\thanks{\footnotemark[2] Average value of two independent measurements; 
PC3=2.80 (SSO) and PC3=2.50 (NTT)}
\thanks{\footnotemark[3] Average value of two independent measurements; 
PC3=2.22 (SSO) and PC3=1.98 (NOT)} 
\thanks{\footnotemark[4] Average value of two independent measurements; 
PC3=1.86 (SSO) and PC3=1.97 (Blanco)}
  \end{list}
\end{table*} 

\clearpage

\setcounter{table}{2}
\begin{table*}
    \caption{continued.}
    \label{obs}
  $$
\begin{tabular}{llll}
   \hline 
   \hline
   \noalign{\smallskip}
DENIS or name &  PC3 &  SpT  &  Notes                 \\
 (1)          &  (2) &  (3)  &  (4)                   \\
   \hline
    \noalign{\smallskip}
J1901391$-$370017 & 1.81 &  M8   & low-gravity \\ 
J1907440$-$282420 & 2.08 & dM9   &           \\ 
J1926005$-$650006 & 2.04 & dM9   &           \\ 
J1934511$-$184134 & 2.01 & dM8.5 &           \\ 
J1935560$-$284634 & 2.33 & dM9.5 & low-gravity \\ 
J1956460$-$774717 & 2.35 & dM9.5 &           \\ 
J2013108$-$124244 & 3.04 & dL1.5 &           \\
J2030412$-$363509 & 1.83 & dM8   &           \\ 
J2045024$-$633206 & 2.23 & dM9.5 &   M9 (6)     \\ 
J2126340$-$314322 & 2.32 & dM9.5 &           \\ 
J2139136$-$352950 & 2.40 & dL0   &           \\ 
J2143510$-$833712 & 2.21 & dM9.5 &           \\ 
J2150133$-$661036 & 2.58 & dL0   &           \\ 
J2150149$-$752035 & 2.91 & dL1   &           \\ 
J2243169$-$593219 & 2.10 & dM9   &  L0 (3)   \\ 
J2308113$-$272200 & 2.97 & dL1.5 &           \\ 
J2322468$-$313323 & 2.90 & dL1   &           \\ 
J2329343$-$540854   & 4.26 & dL3   &           \\ 
J2330226$-$034717 & 2.78 & dL0.5 &  L1 (1)   \\ 
J2345390$+$005514 & 2.10 & dM9   &           \\ 
J2354599$-$185221 & 3.18 & dL2   &           \\ 
GJ 406             & 1.52 & dM6   &  dM6 (12)   \\ 
LHS 2397a           & 2.13 & dM9   &  dM8.5 (12) \\ 
LHS 2924            & 2.25 & dM9.5 &  dM9 (12)  \\
LP 944-20           & 2.12 & dM9   &           \\
VB 8                & 1.65 & dM7   &  dM7 (12)  \\
VB 10               & 1.91\footnotemark[4] & dM8   &  dM8 (12)  \\
DENIS 104814$-$395606 & 2.28 & dM9.5 & M9   (11)          \\ 
DENIS 122815$-$154733 & 10.1 & dL5   & L4.5 (12) \\
DENIS 144137$-$094559 & 2.63 & dL1   & dL1   (12) \\  
2MASS 003615+182112   & 3.71 & dL2.5 & L3.5  (13) \\ 
    \noalign{\smallskip}
    \hline 
   \end{tabular}
 $$
  \begin{list}{}{}
  \item[]
  \end{list}
\end{table*}

\clearpage
\begin{table*}
    \caption{Equivalent widths and molecular band indices}
    \label{indices}
  $$
\begin{tabular}{lllllll}
   \hline 
   \hline
   \noalign{\smallskip}
Name   &  EW H$\alpha$ &  EW NaI(8170--8200) &  TiO & VO &  CrH & FeH  \\
 (1)   &  (2) &  (3) &  (4) &  (5) &  (6) &  (7) \\
   \hline
    \noalign{\smallskip}
DENIS J0000286$-$124514 & -4.5$\pm$-1 & 7.3$\pm$0.4  & 3.23 & 2.40 & 0.99 & 1.04 \\ 
DENIS J0006579$-$643654 & -15$\pm$-2  & 3.0$\pm$0.4  & 3.61 & 2.65 & 0.92 & 0.84 \\ 
DENIS J0014554$-$484417 & $>$-4       & 8.6$\pm$0.8   & 2.29 & 2.06 & 1.47 & 1.53 \\ 
DENIS J0028554$-$192716 & -4$\pm$-1   & 6.6$\pm$0.9  & 2.43 & 2.09 & 1.22 & 1.30 \\ 
DENIS J0031192$-$384035\footnotemark[1] & $>$-7 & 7.8$\pm$0.3  & 2.22 & 2.11 & 1.54 & 1.55 \\ 
DENIS J0031192$-$384035\footnotemark[2] & $>$-2 & 7.5$\pm$0.4  & 2.12 & 2.26 &      & \\ 
DENIS J0050244$-$153818\footnotemark[1] & $>$-3 & 8.8$\pm$0.7 & 2.92 & 2.42 & 1.20 & 1.27 \\
DENIS J0050244$-$153818\footnotemark[2] & $>$-4 & 7.6$\pm$1.2 & 2.87 &      &      &      \\ 
DENIS J0053189$-$363110 & $>$-4       & 6.4$\pm$0.3   & 1.94 & 1.96 & 1.51 & 1.47 \\ 
DENIS J0055005$-$545026 & $>$-8       & 6.8$\pm$0.8   & 3.81 & 2.80 & 0.91 & 0.95 \\ 
DENIS J0116529$-$645557 & $>$-5       & 7.5$\pm$0.8   & 2.90 & 2.24 & 1.23 & 1.28 \\ 
DENIS J0128266$-$554534 & $>$-3       & 6.6$\pm$0.4   & 2.02 & 2.16 & 1.41 & 1.56 \\ 
DENIS J0141582$-$463358 & $>$-3       & 2.1$\pm$0.6   & 2.48 & 2.60 & 0.97 & 0.97 \\ 
DENIS J0147327$-$495448 & -17$\pm$-1  & 7.4$\pm$0.9  & 3.88 & 2.64 & 0.97 & 0.94 \\ 
DENIS J0205294$-$115925 & $>$-3       & 5.2$\pm$0.2   & 1.81 & 1.68 & 1.44 & 1.23 \\ 
DENIS J0206566$-$073519 & -18$\pm$-2  & 8.3$\pm$0.6   & 3.67 & 2.52 & 0.99 & 1.04 \\ 
DENIS J0213371$-$134322 & $>$-5       & 7.6$\pm$0.8   & 3.41 & 2.65 & 0.96 & 0.98 \\ 
DENIS J0224120$-$763320 & -11$\pm$4   & 6.1$\pm$0.9   & 2.63 & 2.27 &      &      \\ 
DENIS J0227102$-$162446 & $>$-4       & 7.2$\pm$0.6   & 1.94 & 1.96 & 1.51 & 1.47 \\ 
DENIS J0230450$-$095305 & $>$-7       & 5.8$\pm$0.7   & 2.30 & 2.01 & 1.40 & 1.41 \\ 
DENIS J0240121$-$530552 & $>$-3       & 5.5$\pm$0.8   & 3.53 & 2.54 & 1.09 & 0.95 \\ 
DENIS J0301488$-$590302 & $>$-5       & 4.5$\pm$0.8   & 4.10 & 2.59 & 0.92 & 0.85 \\ 
DENIS J0314352$-$462341 & --          & 6.5$\pm$0.2   & 2.29 & 2.44 & 1.38 & 1.49 \\ 
DENIS J0325293$-$431229 & $>$-7       & 5.1$\pm$0.8   & 4.04 & 2.89 & 0.95 & 0.96 \\ 
DENIS J0357290$-$441731 & --          & 4.3$\pm$0.3   & 2.80 & 2.87 & 1.03 & 0.97 \\ 
DENIS J0427271$-$112713 & -9.0$\pm$1.2 & 8.6$\pm$0.8  & 4.16 & 2.53 & 0.98 & 0.99 \\ 
DENIS J0428510$-$225322 & -5.4$\pm$0.3 & 7.1$\pm$0.5  & 2.43 & 2.38 & 0.98 & 1.13 \\ 
DENIS J0436360$-$295947 & -8.9$\pm$0.8 & 7.7$\pm$0.5  & 4.29 & 2.66 & 0.98 & 0.92 \\ 
DENIS J0443373$+$000205 & -4.3$\pm$0.5 & 3.6$\pm$0.7  & 4.03 & 2.77 & 0.92 & 0.84 \\ 
DENIS J0529572$-$200300\footnotemark[3] & -10.8$\pm$1.6 & 7.0$\pm$0.8  & 3.46 & 2.56 & 0.99 & 1.04 \\ 
DENIS J0529572$-$200300\footnotemark[4] & -9.7$\pm$2.2 & 7.3$\pm$1.0  & 3.72 & 3.22 &  1.08 & 0.97 \\ 
DENIS J0608528$-$275358 & -3.9$\pm$0.5  & 5.0$\pm$0.7  & 3.97 & 2.93 &  0.94 & 0.80 \\ 
DENIS J0610008$-$472741   & -3.2$\pm$0.5  & 8.8$\pm$0.8  & 3.43 & 2.55 & 1.05 & 1.06 \\ 
DENIS J0620165$-$430010   & -9.0$\pm$0.3  &  7.7$\pm$0.5  & 3.03 & 2.74 & 0.98 & 0.97 \\ 
DENIS J0719317$-$505141 & -12:        & 9.6$\pm$0.8  & 2.42 & 2.50 & 0.64 & 1.02 \\ 
DENIS J0921141$-$210445 & $>$-8       & 9.4$\pm$0.8  & 2.25 & 2.03 & 1.21 & 1.34 \\ 
DENIS J0953213$-$101420 & $>$-2       & 5.4$\pm$0.9   & 3.24 & 2.68 & 0.96 & 0.89 \\ 
DENIS J1004283$-$114648\footnotemark[4] & $>$-10 & -- & 2.36 & 3.29 & 1.05 & 1.18 \\ 
DENIS J1004283$-$114648\footnotemark[5] & -3.3$\pm$0.2 & --           & --  & --  & --  & -- \\
DENIS J1004403$-$131818 &  $>$-2       & 5.1$\pm$0.5   & 1.86 & 1.96 & 1.03 & 1.32 \\ 
DENIS J1019245$-$270717 & $>$-6       &  7.1$\pm$1.0  & 2.59 & 2.31 & 1.09 & 1.05 \\  
DENIS J1115297$-$242934 & -5.8$\pm$0.7 & 7.9$\pm$0.4  & 3.94 & 2.46 & 0.94 & 0.95 \\ 
DENIS J1206501$-$393725 & $>$-4        & 8.6$\pm$0.8 & 2.26 & 2.08 & 1.38 & 1.43 \\ 
DENIS J1216121$-$125731 & $>$-5        & 8.0$\pm$0.7  & 3.78 & 1.85 & 1.11 & 0.92 \\ 
DENIS J1234018$-$112407 & $>$-6        & 8.1$\pm$0.9  & 2.71 & 2.65 & 0.98 & 0.93 \\ 
DENIS J1256569$+$014616 & $>$-5        & 9.0$\pm$0.3  & 1.97 & 2.17 & 1.36 & 1.46 \\ 
DENIS J1359551$-$403456 & $>$-3        & 9.6$\pm$0.4  & 2.45 & 2.32 & 1.19 & 1.34 \\ 
DENIS J1411051$-$791536 & $>$-8        & 6.3$\pm$0.8 & 4.05 & 2.54 & 0.99 & 0.90 \\ 
DENIS J1555256$-$181748 & -23.7$\pm$0.2 & 3.5$\pm$0.1 & 3.46 & 2.56 & 0.87 & 0.71 \\ 
DENIS J1600256$-$192750 & -18.4$\pm$0.2 & 3.6$\pm$0.1 & 3.41 & 2.56 & 0.87 & 0.72 \\ 
DENIS J1602043$-$205043 & -20.0$\pm$0.2 & 4.1$\pm$0.1 & 3.50 & 2.55 & 0.88 & 0.73 \\ 
DENIS J1602553$-$192243 & -17.5$\pm$0.2 & 3.9$\pm$0.1 & 3.41 & 2.56 & 0.88 & 0.74 \\ 
DENIS J1611014$-$192449 & -9.8$\pm$0.4  & 3.9$\pm$0.1 & 3.10 & 2.47 & 0.89 & 0.78 \\ 
DENIS J1611124$-$192737 & -50.1$\pm$0.8 & 2.8$\pm$0.1 & 3.33 & 2.51 & 0.87 & 0.76 \\ 
DENIS J1611296$-$190029 & -12.2$\pm$1.3 & 2.6$\pm$0.2 & 3.66 & 2.63 & 0.83 & 0.71 \\ 
DENIS J1622326$-$120719 & $>$-3        & 7.9$\pm$0.5  & 3.49 & 2.59 & 1.05 & 1.06 \\ 
DENIS J1633131$-$755322 & $>$-5        & 9.3$\pm$0.7 & 3.74 & 2.95 & 1.01 & 1.03 \\ 
DENIS J1703356$-$771520 & -15.6$\pm$2.4 & 1.5$\pm$0.6 & 4.43 & 3.09 & 0.81 & 0.68 \\ 
DENIS J1707252$-$013809 & $>$-6        &  9.1$\pm$1.1  & 2.42 & 2.22 & 1.45 & 1.48 \\ 
DENIS J1716352$-$031542 & -3.3$\pm$0.2 & 2.7$\pm$0.2  & 4.26 & 2.50 & 0.81 & 0.66 \\ 
DENIS J1753452$-$655955 & $>$-3        & 8.0$\pm$0.2  & 1.73 & 2.31 & 1.61 & 1.81 \\ 
    \noalign{\smallskip}
    \hline 
   \end{tabular}
 $$
  \begin{list}{}{}
  \item[]
Column 2: H$\alpha$ equivalent width in \AA .
Column 3: NaI 818.3,819.5 nm doublet equivalent width in \AA .
Column 4: Sum of TiO indices defined in M99.
Column 5: Sum of VO indices defined in M99.
Column 6: CrH index defined in M99. 
Column 7: FeH index defined in M99.
\thanks{\footnotemark[1] Measurements from SSO spectrum} 
\thanks{\footnotemark[2] Measurements from NTT spectrum}
\thanks{\footnotemark[3] Measurements from WHT spectrum}
\thanks{\footnotemark[4] Measurements from NOT spectrum} 
\thanks{\footnotemark[5] Measurements from VLT spectrum}
\thanks{\footnotemark[6] Measurements from CTIO Blanco spectrum} 
  \end{list}
\end{table*}

\clearpage
\setcounter{table}{3}
\begin{table*}
    \caption{continued.}
    \label{indices}
  $$
\begin{tabular}{lllllll}
   \hline 
   \hline
   \noalign{\smallskip}
Name   &  EW H$\alpha$ &  EW NaI(8170--8200) &  TiO & VO &  CrH & FeH  \\
 (1)   &  (2) &  (3) &  (4) &  (5) &  (6) &  (7) \\
   \hline
    \noalign{\smallskip}
DENIS J1901391$-$370017 & $>$-15       & 2.6$\pm$0.6  & 3.92 & 2.68 & 0.86 & 0.72 \\ 
DENIS J1907440$-$282420 & -18.5$\pm$0.6 & 8.7$\pm$0.8 & 3.99 & 2.56 & 0.94 & 0.93 \\ 
DENIS J1926005$-$650006 & -7.5$\pm$1.5 & 5.1$\pm$0.1 & 3.55 & 2.70 & 0.97 & 0.91 \\ 
DENIS J1934511$-$184134 & $>$-10       & 7.2$\pm$0.7  & 3.95 & 2.67 & 0.94 & 0.99 \\ 
DENIS J1935560$-$284634 & 42.1$\pm$1.2 & 1.6$\pm$0.2 & 3.67 & 2.99 & 0.89 & 0.76 \\ 
DENIS J1956460$-$774717 & -4.3$\pm$1.0 & 7.2$\pm$0.1 & 3.76 & 2.66 & 1.06 & 1.09 \\ 
DENIS J2013108$-$124244 & $>$-3        & 9.3$\pm$0.6  & 2.46 & 2.19 & 1.21 & 1.40 \\ 
DENIS J2030412$-$363509 & -2.5$\pm$0.6 & 5.3$\pm$0.4  & 4.00 & 2.89 & 0.92 & 0.80 \\ 
DENIS J2045024$-$633206 & -2.3$\pm$0.5 & 5.9$\pm$0.6  & 3.54 & 2.76 & 0.95 & 0.87 \\ 
DENIS J2126340$-$314322 & -12.9$\pm$0.8 & 9.7$\pm$0.8 & 4.39 & 2.75 & 1.04 & 1.06 \\
DENIS J2139136$-$352950 &  $>$-2        & 8.1$\pm$0.8 & 3.29 & 2.32 & 1.10 & 1.06 \\
DENIS J2143510$-$833712 & $>$-1         & 6.9$\pm$0.6 & 3.71 & 2.54 & 0.90 & 0.84 \\ 
DENIS J2150133$-$661036 & $>$-1         & 6.4$\pm$0.5 & 3.62 & 2.53 & 0.99 & 0.97 \\ 
DENIS J2150149$-$752035 & $>$-2         & 7.3$\pm$0.2 & 2.28 & 2.37 & 1.18 & 1.19 \\ 
DENIS J2243169$-$593219 & $>$-7         & 8.8$\pm$0.3 & 2.92 & 2.35 & 0.99 & 1.02 \\ 
DENIS J2308113$-$272200 & $>$-4         & 9.4$\pm$0.7 & 1.88 & 2.26 & 1.21 & 1.29 \\ 
DENIS J2322468$-$313323 & $>$-4         & 6.4$\pm$0.4  & 2.33 & 2.35 & 1.08 & 1.23 \\ 
DENIS J2329343$-$540854   & --            & 6.9$\pm$0.3  & 2.16 & 2.30 & 1.31 & 1.40 \\ 
DENIS J2330226$-$034717 & $>$-2         & 6.8$\pm$0.4  & 2.64 & 2.37 & 1.20 & 1.22 \\ 
DENIS J2345390$+$005514 & -13.4$\pm$0.7 & 5.9$\pm$0.6 & 3.87 & 2.54 & 0.99 & 0.97 \\ 
DENIS J2354599$-$185221 & $>$-2         & 7.0$\pm$0.4  & 2.21 & 2.27 & 1.26 & 1.48 \\ 
GJ 406                   & -9.4$\pm$0.4 &  9.7$\pm$0.5  &      &      &      &      \\ 
LHS 2397a                & -17.6$\pm$0.2 & 8.0$\pm$0.6 & 3.61 & 2.52 & 1.04 & 1.06 \\
LHS 2924                  & -4.5$\pm$0.3 & 6.1$\pm$0.7 & 3.05 & 2.46 & 1.01 & 1.00 \\
LP 944-20                & -1.5$\pm$0.2 & 7.3$\pm$0.8 & 2.90 & 2.52 &  1.00 & 0.93 \\
VB 8                      & -8.9$\pm$0.3 & 8.9$\pm$0.6 & 3.85 & 2.55 & 0.96 & 0.99 \\
VB 10  \footnotemark[1]                   & -7.1$\pm$0.2 & 7.3$\pm$0.6 & 3.88 & 2.61 & 0.96 & 0.93 \\
VB 10  \footnotemark[6]                   & -8.8$\pm$0.1 & 6.2$\pm$0.3 & 4.01 & 2.71 & 0.95 & 0.91 \\
DENIS J1048147$-$395606 & -2.1$\pm$0.2 & 8.6$\pm$0.6  & 4.03 & 2.75 & 1.05 & 1.07 \\ 
DENIS J1228152$-$154733   & --           & 5.4$\pm$0.3  & 1.73 & 2.33 & 2.07 & 1.78 \\
DENIS J1441373$-$094559 & $>$-7        &  6$\pm$2 & 2.32 & 2.68 & 1.03 & 1.26 \\ %
2MASS 003615+182112      & $>$-2         & 6.8$\pm$0.3 & 2.01 & 2.18 & 1.74 & 1.66 \\
    \noalign{\smallskip}
    \hline 
   \end{tabular}
 $$
  \begin{list}{}{}
  \item[]
  \end{list}
\end{table*}

\clearpage
\begin{table*}
    \caption{Absolute magnitudes and spectrophotometric distances for confirmed 
nearby dwarfs not known to be binaries}
    \label{dist}
  $$
\begin{tabular}{lllll}
   \hline 
   \hline
   \noalign{\smallskip}
DENIS name  &   SpT  &  M$_J$  &  Distance (pc) &  Distance error    \\
 (1)        &   (2)  &  (3)    &  (4)           &  (5)               \\                       
   \hline
    \noalign{\smallskip}
J0000286$-$124514 & dM9.5 & 11.65 & 19.2 & 3.6  \\ 
J0006579$-$643654 & dL0   & 11.83 & 20.9 & 3.8  \\ 
J0014554$-$484417 & dL2.5 & 12.74 & 17.7 & 3.1  \\ 
J0028554$-$192716 & dL0.5 & 12.01 & 24.4 & 5.4  \\ 
J0031192$-$384035 & dL2.5 & 12.74 & 18.7 & 3.4  \\ 
J0050244$-$153818 & dL0.5 & 12.01 & 21.4 & 3.9  \\ 
J0053189$-$363110 & dL2.5 & 12.74 & 19.7 & 3.8  \\ 
J0055005$-$545026 & dM8.5 & 11.28 & 31.0 & 7.1  \\ 
J0116529$-$645557 & dL1   & 12.19 & 28.1 & 5.2  \\ 
J0128266$-$554534 & dL1   & 12.19 & 21.0 & 4.1  \\ 
J0206566$-$073519 & dM8.5 & 11.28 & 40.9 & 7.7  \\ 
J0213371$-$134322 & dM9   & 11.47 & 36.5 & 7.7  \\ 
J0224120$-$763320 & dL0   & 11.83 & 50.8 & 11.2 \\ 
J0227102$-$162446 & dL0   & 11.83 & 23.3 &  4.5 \\ 
J0230450$-$095305 & dL0   & 11.83 & 37.2 &  8.2 \\ 
J0240121$-$530552 & dM9.5 & 11.65 & 34.7 &  6.4 \\ 
J0301488$-$590302 & dM9   & 11.47 & 24.7 &  4.3 \\ 
J0314352$-$462341 & dL2   & 12.56 & 28.9 &  6.0 \\ 
J0325293$-$431229 & dM8.5 & 11.28 & 37.6 &  6.9 \\ 
J0427271$-$112713 & dM7   & 10.92 & 33.3 &  5.8 \\ 
J0428510$-$225322 & dL0.5 & 12.01 & 19.4 &  3.4 \\ 
J0436360$-$295947 & dM8   & 11.10 & 55.2 & 11.2 \\ 
J0443373$+$000205 & dM9.5 & 11.65 & 15.0 &  2.8 \\ 
J0529572$-$200300 & dM9   & 11.47 & 39.9 &  6.0 \\ 
J0608528$-$275358 & dM9.5 & 11.65 & 25.5 &  4.6 \\ 
J0610008$-$472741  & dM8.5 & 11.28 & 43.2 &  8.5 \\ 
J0620165$-$430010  & dM8   & 11.10 & 59.1 & 11.7 \\ 
J0719317$-$505141 & dL1   & 12.19 & 23.0 &  4.1 \\ 
J0921141$-$210445 & dL3   & 12.92 &  9.7 &  1.7 \\ 
J0953213$-$101420 & dL0   & 11.83 & 21.8 &  3.8 \\ 
J1004283$-$114648 & dM8   & 11.10 & 56.2 & 11.6 \\
J1004403$-$131818 & dL0   & 11.83 & 36.8 &  9.2 \\ 
J1019245$-$270717 & dL0.5 & 12.01 & 20.5 &  3.6 \\ 
J1115297$-$242934 & dM8   & 11.10 & 28.5 &  4.9 \\ 
J1206501$-$393725 & dL2   & 12.56 & 22.4 &  4.1 \\ 
J1216121$-$125731 & dM8   & 11.10 & 63.6 & 16.4 \\ 
J1234018$-$112407 & dM9.5 & 11.65 & 39.5 &  7.8 \\ 
J1256569$+$014616 & dL1.5 & 12.38 & 25.8 &  4.6 \\ 
J1359551$-$403456 & dL2   & 12.56 & 17.6 &  3.2 \\ 
J1411051$-$791536 & dM8.5 & 11.28 & 23.4 &  4.1 \\ 
J1622326$-$120719 & dM9.5 & 11.65 & 22.0 &  3.8 \\ 
J1633131$-$755322 & dM9.5 & 11.65 & 19.5 &  3.3 \\ 
J1707252$-$013809 & dL0.5 & 11.47 & 54.8 & 10.1 \\ 
J1753452$-$655955 & dL2   & 12.56 & 21.4 &  3.9 \\ 
    \noalign{\smallskip}
    \hline 
   \end{tabular}
 $$
  \begin{list}{}{}
  \item[]
Column 1: DENIS name.
Column 2: Spectral type adopted in this paper. 
Column 3: Absolute J-band magnitude estimated from
the absolute magnitude vs. $I-J$ color relationship 
given in Phan-Bao et al (2008). 
Column 4: Spectrophometric distance in parsecs. 
  \end{list}
\end{table*}

\clearpage
\setcounter{table}{4}
\begin{table*}
    \caption{continued.}
    \label{dist}
  $$
\begin{tabular}{lllll}
   \hline 
   \hline
   \noalign{\smallskip}
DENIS name  &   SpT  &  M$_J$  &  Distance (pc) &  Distance error    \\
 (1)        &   (2)  &  (3)    &  (4)           &  (5)               \\                       
   \hline
    \noalign{\smallskip}
J1907440$-$282420 & dM9   & 11.47 & 37.7 &  7.3 \\ 
J1926005$-$650006 & dM9   & 11.47 & 41.4 &  8.0 \\ 
J1934511$-$184134 & dM8.5 & 11.28 & 39.7 &  7.5 \\ 
J1956460$-$774717 & dM9.5 & 11.65 & 32.1 &  6.1 \\ 
J2013108$-$124244 & dL1.5 & 12.38 & 26.8 &  5.6 \\
J2030412$-$363509 & dM8   & 11.10 & 43.8 &  7.9 \\ 
J2045024$-$633206 & dM9.5 & 11.65 & 15.9 &  3.1 \\ 
J2126340$-$314322 & dM9.5 & 11.65 & 20.4 &  4.0 \\ 
J2139136$-$352950 & dL0   & 11.83 & 33.7 &  6.4 \\ 
J2143510$-$833712 & dM9.5 & 11.65 & 20.4 &  3.4 \\ 
J2150133$-$661036 & dL0   & 11.83 & 23.4 &  4.1 \\ 
J2150149$-$752035 & dL1   & 12.19 & 22.3 &  4.0 \\ 
J2243169$-$593219 & dM9   & 11.47 & 33.3 &  6.1 \\ 
J2308113$-$272200 & dL1.5 & 12.38 & 27.6 &  6.0 \\ 
J2322468$-$313323 & dL1   & 12.19 & 18.7 &  3.7 \\ 
J2329343$-$540854  & dL3   & 12.92 & 25.6 &  4.9 \\ 
J2330226$-$034717 & dL0.5 & 12.01 & 30.2 &  6.3 \\ 
J2345390$+$005514 & dM9   & 11.47 & 28.2 &  5.6 \\ 
J2354599$-$185221 & dL2   & 12.56 & 21.5 &  3.9 \\ 
    \noalign{\smallskip}
    \hline 
   \end{tabular}
 $$
  \begin{list}{}{}
  \item[]
  \end{list}
\end{table*}

\begin{center}
\begin{table}
\caption{{\it Spitzer} observation log}
\label{spitzer_obs}
\begin{tabular}{lllll}\hline\hline
Object & Program ID & P.I. & Obs. Date & Instrument \\
\hline
J0141582-463358 & 30540 & Houck & 2006-08-12 & IRAC1-4 \\
J0141582-463358 & 284   & Cruz  & 2007-07-16 & MIPS1 \\
J1611296-190029 & 20103 & Hillenbrand & 2005-08-24 & IRAC1-4 \\
J1611296-190029 & 20103 & Hillenbrand & 2006-04-06 & MIPS1 \\
J1901391-370017 & 6     & Fazio & 2004-04-20 & IRAC1-4 \\
J1901391-370017 & 6     & Fazio & 2004-04-11 & MIPS1 \\
\hline
\end{tabular}
\end{table}
\end{center}

\begin{center}
\begin{table}
\caption{{\it Spitzer} mid-IR photometry}
\label{spitzer}
\begin{tabular}{llllll}\hline\hline
Object    & 3.6~$\mu$m & 4.5~$\mu$m & 5.8~$\mu$m & 8.0~$\mu$m & 24~$\mu$m  \\
          & [mJy]      & [mJy]      & [mJy]      & [mJy]      & [mJy]      \\
\hline
J0141582-463358 & 3.29  & 2.47 & 1.86 & 1.61 & 0.20 \\
J1611296-190029 & 4.81  & 3.40 & 2.34 & 1.46 & $<$0.25 \\
J1901391-370017 & 6.48 & 5.02 & 3.53 & 2.07 & $<$2.27 \\
\hline
\end{tabular}
\end{table}
\end{center}

\end{document}